\documentclass[final,3p,times]{elsarticle}

\usepackage{amssymb}
\usepackage{amsmath}
\usepackage[normalem]{ulem}
\usepackage{xcolor}

\journal{Annals of Physics}

\begin{document}

\begin{frontmatter}



\title{Scale-dependent theory of the disordered electron liquid}

\author[1,2]{A. M. Finkel'stein}
\author[3]{G. Schwiete}

\affiliation[1]{organization={Department of Condensed Matter Physics, the Weizmann Institute of Science},
            city={Rehovot},
            postcode={76100}, 
            country={Israel}}
            
\affiliation[2]{organization={Department of Physics and Astronomy, Texas A\&M University},
            city={College Station},
            postcode={77843-4242}, 
            state={Texas},
            country={USA}}
            
\affiliation[3]{organization={Department of Physics and Astronomy, The University of Alabama},
 %           addressline={}, 
            city={Tuscaloosa},
            postcode={35487}, 
            state={Alabama},
            country={USA}}

\begin{abstract}
We review the scaling theory of disordered itinerant electrons with $e$-$e$ interactions. We first show how to adjust the microscopic Fermi-liquid theory to the presence of disorder. Then we describe the non-linear sigma model (NLSM) with interactions (Finkel'stein's model). This model is closely connected to the Fermi liquid, but is more generally applicable, since it can also be viewed as  a minimal effective functional describing disordered interacting electrons. Our discussion emphasizes the general structure of the theory, and the connection of the scaling parameters to conservation laws. We then move on to discuss the metal-insulator transition (MIT) in the disordered electron liquid in two and three dimensions. Generally speaking, this MIT is a non-trivial example of a quantum phase transition. The NLSM approach allows to identify the dynamical exponent connecting the spatial and energy scales, which is central for the description of the kinetic and thermodynamic behavior of the system in the critical region of the MIT in three dimensions. In two dimensions, the system can be discussed in terms of a flow in the disorder-interaction phase plane, which is controlled by a fixed point. We demonstrate that the two-parameter RG-equations accurately describe electrons in Si-MOSFETs including the observed non-monotonic behavior of the resistance and its strong drop at low temperatures. The theory can also be applied to systems with an attractive interaction in the Cooper channel, where it describes the suppression of superconductivity in disordered amorphous films. We extend our discussion to heat transport in the two-dimensional electron liquid. Similar to the electric conductivity, the thermal conductivity also acquires logarithmic corrections induced by the interplay of the electron interaction and disorder. The resulting thermal conductivity can be calculated in the NLSM formalism after introducing so-called gravitational potentials. It may be concluded that the combined effect of disorder and $e$-$e$ interactions determines all aspects of the physics of itinerant electrons in the presence of disorder.  
\end{abstract}


\end{frontmatter}

\section{Introduction}

Landau's original Fermi liquid theory has been formulated for clean interacting systems. The limits of applicability of this theory are reached when the production of multiple electron-hole pairs becomes too effective and quasi-particles are no-longer well-defined. Elastic impurity scattering by itself does not generate electron-hole excitations and, therefore, some elements of the Fermi-liquid description carry over to disordered electron systems. The focus, however, shifts from quasiparticles to diffusing electron-hole pairs. Such pairs are described by a singular diffusion propagator $1/(Dq^2-i\omega)$, also known as the "diffuson". For temperatures smaller than the elastic scattering rate, $T\lesssim 1/\tau _{el}$, diffusion modes rather than quasi-particles determine the physics on large spatial scales. Furthermore, the interaction amplitudes acquire non-analytic corrections in the presence of disorder. Conversely, the resistivity - which is a measure of the effective disorder strength - also acquires corrections which depend on the value of the interaction amplitudes.

In dimension $d=2$, \emph{all} corrections both to the diffusion coefficient (electric conductivity) and the interaction amplitudes are proportional to $\ln 1/T\tau_{el}$, i.e., logarithmically divergent at low temperatures. In higher dimensions, the problem becomes logarithmic near the MIT. The divergent corrections signal that there is a need for a scale-dependent description of the system. The scheme most suitable for the analysis of the singular corrections arising in the system of diffusing electrons is the Renormalization Group (RG). In the RG-analysis, the scattering rate $1/\tau _{el}$ acts as a high-energy cutoff, because only states with energy smaller than $1/\tau _{el}$ are relevant in the diffusive regime, while the temperature $T$ enters as a low-energy cutoff. The RG-scheme is able to account for both effects, the disorder and the $e$-$e$ interactions, in a consistent fashion, producing as an output a system of \emph{coupled} equations. 

Since the RG-analysis requires a synthesis of very different phenomena, the original Hamiltonian written in terms of the fermionic fields is not appropriate for this task. Instead, the RG-analysis of the disordered electron liquid can be performed using a mapping of the original fermion problem onto the effective matrix functional, the NLSM. This treatment was originally developed for non-interacting electrons, but it was possible to extend it by introducing additional terms that describe electron-electron interactions including the long-range Coulomb interaction (the Finkel'stein model). Although the derivation of the latter functional has been performed under certain constraints, the result is a minimal functional that incorporates all essential degrees of freedom and symmetries. As such, it is not a model, in a similar sense as Landau's Fermi-liquid theory is not a model. Treated within the Gaussian approximation (i.e., keeping only the quadratic expansion of the matrices with respect to their equilibrium values), the Finkel'stein model reproduces the Fermi-liquid description of the electron liquid. The conservation laws (number of particles, spin, and energy) introduce strong limitations on the form of the correlation functions which can be calculated with the use of the NLSM with electron interactions. The dynamic exponent connecting the spatial and energy scales, which is central for the description of the critical region of the MIT, is determined by one of the RG equations. With this equation at hand, the technique of the matrix NLSM allows to understand the critical behavior of the MIT in dimension higher than two. This includes thermodynamic quantities as well as kinetic ones. In two dimensions, the RG-flow captures both quantitatively and qualitatively the physics of the disordered electron liquid. The experimental data obtained for $2d$ electrons in MOSFETs can be presented as a flow in the interaction-disorder phase plane and provides strong support of the scaling theory. The theory can be extended to the description of amorphous superconducting films, where disorder can lead to a suppression of the superconducting transition temperature. Similar to the electric conductivity, the thermal conductivity also acquires pronounced temperature corrections at low temperatures, which can be analyzed by extending the NLSM to include so-called gravitational potentials. The scaling theory for the description of the disordered electrons has been introduced originally as an almost heuristic hypothesis by extending the original ideas of Thouless. The technique of the matrix NLSM allows to put the scaling theory of the disordered electron liquid on firm grounds. 

This manuscript is structured as follows. In Sec.~\ref{sec:FL}, we introduce elements of the Fermi-liquid description of the disordered electron liquid. We discuss the density-density and spin-spin correlation functions and their relation to the transport coefficients for charge and spin. Sec.~\ref{sec:nlsm} is devoted to the NLSM approach for interacting electron systems, a powerful field-theoretical description of the disordered electron liquid. The NLSM can be subjected to an RG treatment in order to analyze non-analytic corrections to transport coefficients and thermodynamic quantities that arise at low temperatures. The structure of Fermi-liquid theory is preserved during the RG procedure, albeit with scale and temperature-dependent parameters. The resulting scaling theory of the MIT in three dimensions is the subject of Sec.~\ref{sec:scaling}. Section~\ref{sec:SC} is concerned with the application of the RG-analysis to amorphous superconducting films. The MIT in two-dimensional disordered systems is reviewed in Sec.~\ref{sec:2dMIT}, with particular emphasis on the comparison between theory and experiment. In Sec.~\ref{sec:Kappa}, we generalize the RG approach to the thermal conductivity, and discuss the origin of the violation of the Wiedemann-Franz law in the disordered electron liquid. We finally conclude with a brief summary in Sec.~\ref{sec:Summary}. 

This review extends and updates the material discussed in Ref.~\cite{Finkelstein_Anderson50}. A compact presentation of the subjects covered in this review can be found in Ref.~\cite{Finkelstein22}.

\section{Fermi-Liquid in the presence of disorder: $\mathbf{1/\protect\tau _{el}>T,\omega,Dk^2}$}
\label{sec:FL}

In Landau's original microscopic theory of the clean Fermi-liquid \cite{Pitaevskii,Nozieres}, the term $v_{F}\mathbf{nk}/(\omega -$ $v_{F}\mathbf{nk})$ is used as the propagator of an electron-hole pair, see Ch. 2, \S 17 in Ref.~\cite{Pitaevskii}. This expression describes propagation of the pair along the direction $\mathbf{n}$, when the momentum difference of the two quasiparticles is $\mathbf{k}$, and the frequency difference is $\omega $. In order to obtain this propagator, one needs to integrate out the energy variable $\xi=\epsilon(\mathbf{p})-\mu$. The product $v_{F}\mathbf{nk}$ originates from the energy difference of the constituents of the pair, $\delta \epsilon _{k}(\mathbf{p)}=\epsilon (\mathbf{p}+\mathbf{k})-\epsilon (\mathbf{p})\approx v_{F}\mathbf{nk}$ where $\mathbf{n}=\mathbf{p}/p$. The singularity at small momenta and frequencies of the electron-hole pair propagator carries over to the two-particle vertex function $\Gamma (\omega ,\mathbf{k)}$ since this function describes, among other processes, rescattering of electron-hole pairs. The propagator $v_{F}\mathbf{nk}/(\omega -$ $v_{F}\mathbf{nk})$ may be rewritten as the sum of a static and a dynamical part: $[-1+\omega /(\omega-v_{F}\mathbf{nk})]$. In fact, it is more convenient to keep explicitly only the \emph{dynamical} part of this propagator, $\omega /(\omega -v_{F}\mathbf{nk})$, and to delegate the static part (equal to $-1$) to the amplitude of the electron-electron ($e$-$e)$ interaction. The resulting \emph{static} amplitude is denoted as $\Gamma ^{k}$ [N.1] (to keep the discussion focused we delegate some notes to a separate section following the main text).

\begin{figure}[h]
\begin{flushright}\begin{minipage}{1\textwidth}  \centering
        \includegraphics[width=0.9\textwidth,clip=]{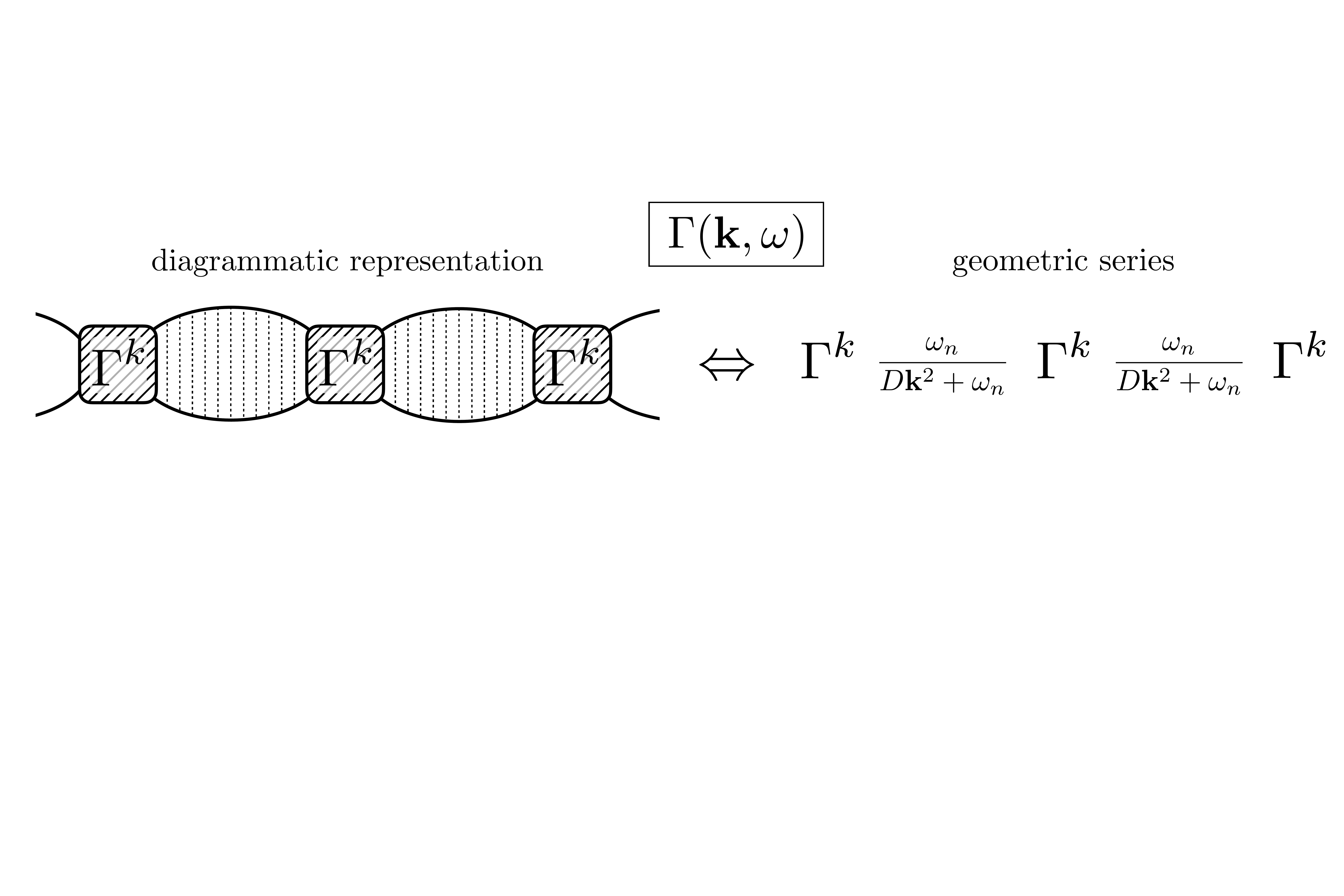}
        \caption{\small Disordered Fermi-liquid: geometric series leading to Eq.~(\ref{ladder}). Dashed lines symbolize the impurity scattering
        after averaging over a random disorder realization. In the framework of Fermi liquid theory only ladder-like processes are kept.} \label{fig:ladder}
    \end{minipage}
\end{flushright}
\end{figure}

For large times, $t\gg \tau _{el}$, electrons have undergone many collisions with impurities. Then, after averaging over the directions $\mathbf{n}$, the dynamical part of the electron-hole pair propagator acquires a diffusive form 
\begin{equation}
\frac{\omega}{\omega-v_{F}\mathbf{nk}}\Longrightarrow \frac{\omega _{n}}{\omega _{n}+iv_{F}\mathbf{nk}}\Longrightarrow \frac{\omega _{n}}{\omega _{n}+Dk^{2}},\label{diffusive}
\end{equation}
where $D=v_{F}^{2}\tau _{el}/d$ is the diffusion coefficient for the spatial dimension $d$ [N.2]; it is assumed that $1/\tau_{el}>D{\bf k}^2,\omega, T$. In $D$, both $v_{F}$ and $\tau_{el}$ incorporate the Fermi-liquid renormalizations. In view of potential complications induced by the singular denominator,
the analytical continuation of $\omega$ to the imaginary (Matsubara) axis was performed here. The right term in Eq. (\ref{diffusive}) describes the free diffusion of electron-hole pairs between acts of rescattering  induced by $e$-$e$ interactions as illustrated in Fig.~\ref{fig:ladder}. 

In clean Fermi-liquids, different angular harmonics $\Gamma _{l}^{k}$ of the interaction amplitudes are coupled. On the contrary, in the presence of disorder only the zeroth harmonic of the
interaction amplitude, $\Gamma _{l=0}^{k}$, is relevant for the rescattering of diffusing electron-hole pairs [N.3]. All other harmonics with $l\neq0$ are eliminated by random impurity scattering. As a result, the calculation of $\Gamma (\mathbf{k},\omega \mathbf{)}$ reduces to a simple summation of a geometric series. 
The two-particle amplitude $\Gamma _{l=0}^{k}$ can be split into parts which can be classified according to their spin structure:
\begin{eqnarray}
\nu a^{2}{\Gamma _{l=0}^{k}}_{\alpha _{3}\,\alpha _{4}}^{\alpha
_{1}\alpha _{2}} &=&\frac{1}{2}\left[2\widetilde{\Gamma }_{\rho}\delta _{\alpha
_{1},\alpha _{3}}\delta _{\alpha _{2},\alpha _{4}}+\Gamma _{\sigma}
\vec{\sigma}_{\alpha _{1},\alpha _{3}}\vec{\sigma}_{\alpha _{2},\alpha _{4}}\right].\label{Gamma}
\end{eqnarray}
Here, the singlet-channel amplitude $\widetilde{\Gamma }_{\rho }$ controls propagation of the particle-number density $\rho(\mathbf{k},\omega)$ in the singlet channel, $S=0$, while $\Gamma _{\sigma }$ controls the spin density, i.e., the triplet channel, $S=1$ [N.4]. In Eq.~(\ref{Gamma}), $\nu $ is the single-particle density of states per one spin component at energy $\epsilon _{F}$, and the factor $a^2$ describes the effect of the so-called wave function renormalization of the interaction amplitudes which originates from scales shorter than the mean free path. Renormalizations originating from larger spatial scales as well as energies smaller than $1/\tau_{el}$ are described by the RG, and they will be discussed below.

\begin{figure}[h]
\begin{flushright}\begin{minipage}{1\textwidth}  \centering
        \includegraphics[width=0.82\textwidth,clip=]{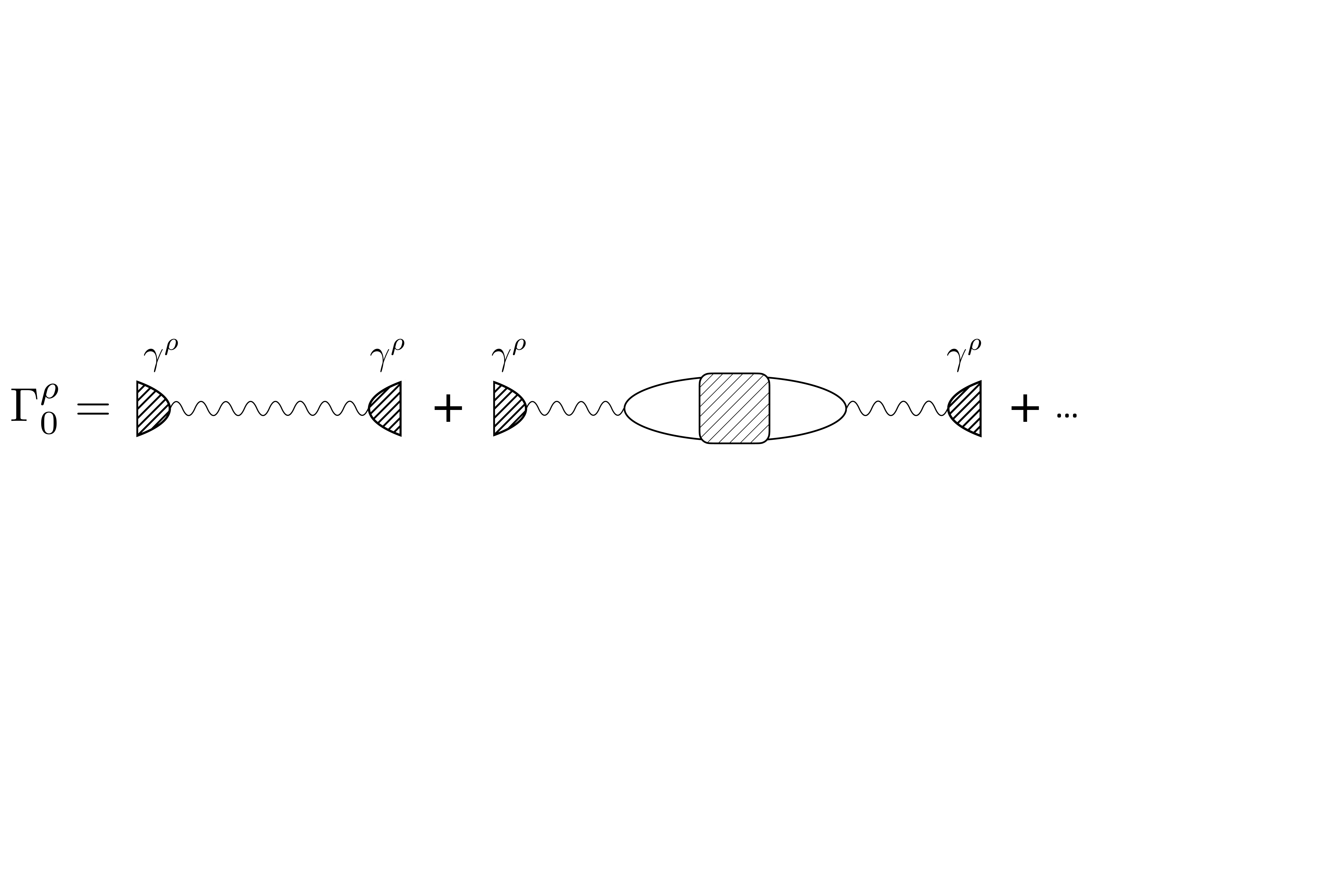}
        \caption{\small The screened Coulomb interaction. The triangular vertices $\gamma^{\rho}$ are attached to the ending points of the interaction line. The ellipse with shaded square stands for the polarization operator. } \label{fig:Coulombladder}
    \end{minipage}
\end{flushright}
\end{figure}

A key point in studies of the true electron liquid, i.e., a quantum liquid with \emph{charge-carrying fermions} is the long-range character of the Coulomb interaction which requires special care. With this in mind, one has to identify terms in the interaction amplitude $\widetilde{\Gamma }_{\rho }$ that carry information about the singular behavior of the Coulomb interaction at small momenta. (Graphically, these terms can be disconnected
by cutting a single Coulomb interaction line, see Fig.~\ref{fig:Coulombladder}.) 
The resulting amplitude of this part of the $e$-$e$ interaction will be denoted as $\Gamma _{\rho }^{0}$, while the remaining interaction amplitude, which is irreducible with respect to the Coulomb interaction, will be denoted as $\Gamma _{\rho }$. The latter part describes the short-ranged part of the $e$-$e$ interaction in the singlet channel that is not sensitive to the Coulomb singularity at small momenta. As a result, the amplitude $\widetilde{\Gamma }_{\rho }$ has been split into two terms, $\widetilde{\Gamma }_{\rho }=\Gamma_{\rho }^{0}+\Gamma _{\rho }$. 

To include dynamics in the $\rho$- and $\sigma$-density channels, one has to consider a ladder of either $\Gamma_{\rho }$- or $\Gamma _{\sigma }$-amplitudes alternating with dynamical electron-hole propagators. The resulting amplitudes $\Gamma_{\rho }(\mathbf{k},\omega )$ and $\Gamma _{\sigma }(\mathbf{k},\omega )$ acquire the form:
\begin{equation}
\Gamma _{\alpha }(\mathbf{k},\omega )=\Gamma _{\alpha }\frac{Dk^{2}+\omega
_{n}}{Dk^{2}+(1-\Gamma _{\alpha })\omega _{n}}~,\hspace{0.5cm}\alpha =\rho
,\sigma .  \label{ladder}
\end{equation}
The position of the diffusion pole in $\Gamma _{\rho,\sigma }(k,\omega )$ shifts as a result of rescattering induced by the interaction amplitudes. This shift is the origin of the renormalization of the diffusion coefficients that is different in the $\rho$- and $\sigma$ channels. The amplitudes $\Gamma_{\rho,\sigma}$ are related to the standard Landau parameters of Fermi liquid theory as $1-\Gamma_{\alpha}=1/(1+F_{0}^{\alpha})$.

In the analysis of the density-density correlation functions for a charged liquid, one has to follow the same procedure as discussed above for the interaction amplitude in the $\rho$-channel. The focus of attention has to be directed toward the polarization operator $\Pi (k,\omega _{n})$, 
which is the density-density correlation function \emph{irreducible} with respect to the Coulomb interaction. 
\begin{figure}[h]
\begin{flushright}\begin{minipage}{1\textwidth}  \centering
        \includegraphics[width=0.8\textwidth,clip=]{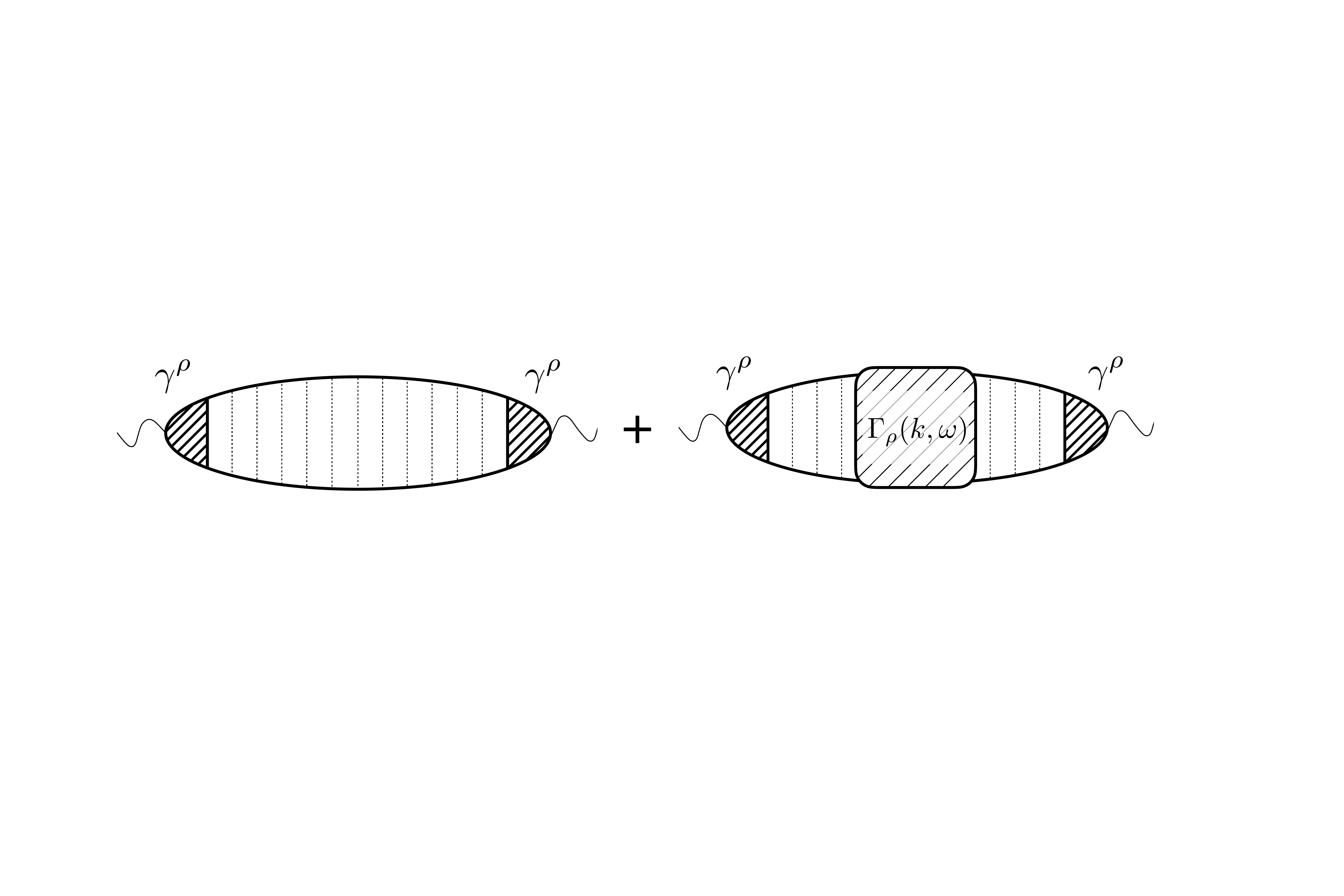}
        \caption{\small Disordered Fermi-liquid: dynamical part of the polarization operator $\Pi (k,\omega _{n})$. \label{fig:polarization_operator}}
    \end{minipage}
\end{flushright}
\end{figure}
The polarization operator $\Pi(k,\omega_n)$ can be written as the sum of a static and a dynamical part, as described for the interaction amplitude $\Gamma_{\alpha}(k,\omega _{n}\mathbf{)}$ above: the dynamical part of the polarization operator contains two ``triangle'' vertices $\gamma ^{\rho }=1-\Gamma_\rho$ separated by a ladder of dynamical electron-hole pair propagators, see Fig.~\ref{fig:polarization_operator}. The left and right vertices $\gamma^\rho$ do not contain the dynamical sections. The static part and the dynamical part of the polarization operator can be combined as follows
\begin{subequations}
\begin{align}
& \Pi (k,\omega _{n})=\Pi _{st}-2\nu \left( \gamma ^{\rho }\right) ^{2}\left[
\frac{\omega _{n}}{Dk^{2}+(1-\Gamma _{\rho })\omega _{n}}\right]  \label{up}
\\
& \Longrightarrow \Pi _{st}\frac{Dk^{2}}{Dk^{2}+(1-\Gamma _{\rho })\omega
_{n}}\;.  \label{bottom}
\end{align}
\end{subequations}
The polarization operator vanishes in the limit $k\rightarrow 0$ at finite $\omega_n\neq 0$, as required for the correlation function of a conserved quantity.

Introducing the diffusion coefficient $D_\rho$ of the particle number density $\rho$, one may rewrite the polarization operator in Eq.~\eqref{bottom} in a diffusive form
\begin{equation}
\Pi (k,\omega _{n})=\Pi _{st}\frac{D_{\rho }k^{2}}{D_{\rho }k^{2}+\omega _{n}
}\;\qquad D_{\rho }=\frac{D}{1-\Gamma _{\rho }}.
\end{equation}
The polarization operator $\Pi(k,\omega_n)$ can be related to the current-current correlation function through the continuity equation $\partial \rho /\partial t+\mathbf{\nabla}\cdot\mathbf{j}=0$. Then, with the help of the Kubo formula, one obtains the Fermi-liquid expression for the electric
conductivity ($e$ is electron charge):
\begin{eqnarray}
\frac{\sigma _{charge}}{e^{2}} &=&\lim_{k\rightarrow 0}\frac{\omega _{n}}{
k^{2}}\Pi (k,\omega _{n})=  \frac{\partial n}{\partial \mu }D_{\rho }~=2\nu D.
\label{eqn:Einsteincharge}
\end{eqnarray}
This is the Einstein relation for the electric conductivity $\sigma \equiv \sigma _{charge}$. It is noteworthy that the electric conductivity is not determined by the na\"ive expression $(\partial n/\partial \mu) D$, but rather by the product $(\partial n/\partial \mu)D_\rho$. This point is very important. The compressibility $\partial n/\partial \mu=2\nu (1-\Gamma_\rho)$ of the dilute two-dimensional electron gas is often negative in heterostructures [N.5]. However, the negative renormalizations of $\partial n/\partial \mu$ and $D_\rho$ cancel when calculating the conductivity, which is unquestionably positive. 

It remains to comment why the $\emph{irreducible}$ part of the density-density correlation function, $\Pi (k,\omega _{n})$, rather than the full correlation function, has been used to find the conductivity, see Ref.~\cite{AF1983}. The terms that can be disconnected by cutting a single Coulomb line describe the rearrangement of the system in response to the external perturbation. The point here is that conductivity is a response to a resulting field acting on a charged electron (the arguments go back to Ref.~\cite{Nozieres}). This resulting field is measured by a voltmeter as the difference in the electro-chemical potential. Thus, by keeping only terms that are irreducible with respect to the Coulomb interaction the rearrangement has been taken into account, and one can obtain the electric current in response to the actual field.

\begin{figure}[h]
\begin{flushright}\begin{minipage}{1\textwidth}  \centering
        \includegraphics[width=0.8\textwidth,clip=]{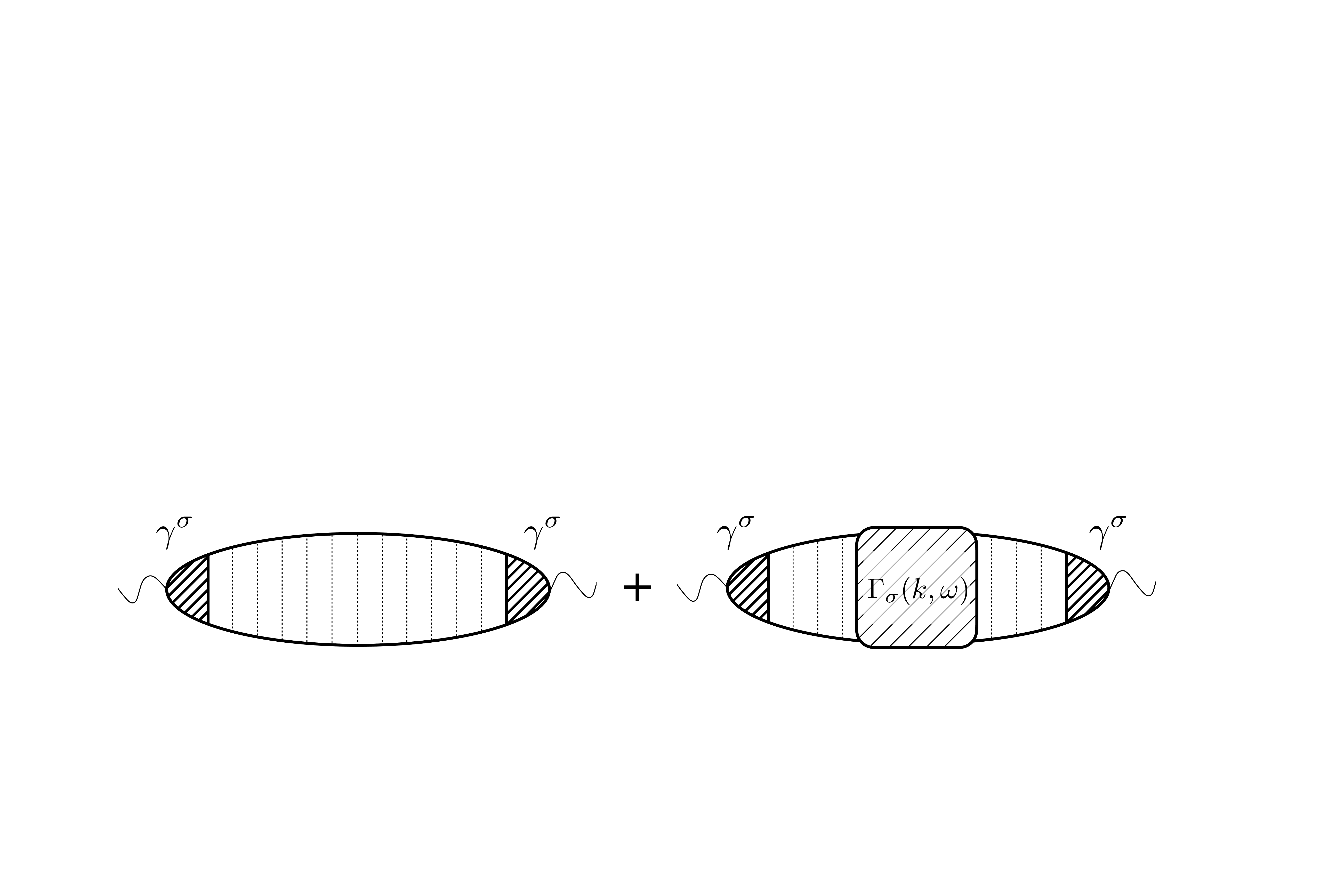}
        \caption{\small Disordered Fermi-liquid: dynamical part of the spin-density
correlation function $\chi _{s}(k,\omega _{n})$. \label{fig:spindensity}}
    \end{minipage}
\end{flushright}
\end{figure}

In the case when the total spin of conducting electrons is conserved, the scheme of Ref.~\cite{AF1983} outlined above can straightforwardly be applied for the analysis of the spin-density correlation function~\cite{AF1984,CastellaniSpin}. In this case, the external vertices contain a spin operator $\sigma ^{x}/2$ that corresponds to a probing magnetic field directed along the $x $-axis. These vertices are renormalized by the $e$-$e$ interactions, and
the corresponding renormalization factor is denoted below as $\gamma^{\sigma }$. In spite of this modification, all formulas for the spin-density correlation function $\chi _{s}^{xx}(k,\omega _{n})$ are similar to those obtained in the case of $\Pi (k,\omega _{n})$. The only change is the substitution of $\Gamma _{\rho }$ by $\Gamma_{\sigma }$ in the above expressions. Correspondingly, one should replace $\gamma^\rho\rightarrow \gamma^\sigma$ and $\Gamma_\rho\rightarrow \Gamma_\sigma$ in Fig.~\ref{fig:polarization_operator}, as illustrated in Fig.~\ref{fig:spindensity}. 

The static part of the spin-density correlation function $\chi_s^{xx}(k,\omega)$ is the spin susceptibility $\chi_s$, just like the static part of the polarization operator is the susceptibility $\partial n/\partial\mu$ (closely related to the compressibility). As a result of the Fermi liquid corrections, $\chi_{s}\equiv\chi _{s}^{xx}(k\rightarrow 0,\omega _{n}=0)$ is modified by the factor $(1-\Gamma _{\sigma })$ compared to $\chi_s^0=(g_L^0\mu _{B}/2)^{2}2\nu$ for the clean Fermi \emph{gas}, 
so that $\chi _{s}=\chi _{s}^{0}(1-\Gamma _{\sigma })$. In the standard Fermi-liquid notation, $1-\Gamma _{\sigma }=1/(1+F_{0}^{\sigma })$. Usually, $F_{0}^{\sigma }$ is negative. Then, $(1-\Gamma _{\sigma })\ $ describes the Stoner enhancement of the spin susceptibility as well as the suppression of the spin-diffusion coefficient $D_{\sigma }=D/(1-\Gamma _{\sigma })$
\begin{eqnarray}
\chi _{s}^{xx}(k,\omega ) =\chi _{s}^{0}(1-\Gamma _{\sigma })\ \frac{Dk^{2}
}{Dk^{2}+(1-\Gamma _{\sigma })\;\omega _{n}} =\chi _{s}\ \frac{D_{\sigma }k^{2}}{\omega _{n}+D_{\sigma }k^{2}}.
\label{spinsuscept}
\end{eqnarray}
The Einstein relation for the spin conductivity reads as
\begin{eqnarray}
\frac{\sigma _{spin}}{(g_L^0\mu _{B}/2)^{2}} &=&\frac{1}{(g_L^0\mu _{B}/2)^{2}}
\lim_{k\rightarrow 0}\frac{\omega _{n}}{k^{2}}\chi _{s}^{xx}(k,\omega _{n})
=2\nu (1-\Gamma _{\sigma })D_{\sigma }=2\nu D.  \label{eqn:Einsteinspin}
\end{eqnarray}
Taken together, Eqs.~(\ref{eqn:Einsteincharge}) and (\ref{eqn:Einsteinspin}) reflect the fact that both the charge and the spin are carried by the same
particles.

\textit{Conclusion}: The theory of the disordered Fermi-liquid focuses on diffusing electron-hole pairs. For temperatures (frequencies) smaller than the elastic scattering rate, $T\lesssim 1/\tau _{el}$, diffusion modes and not quasi-particles determine the physics on large spatial scales. The conservation of the particle-number and the conservation of spin dictate the possible form of the corresponding correlation functions. The theory contains two parameters, $\Gamma _{\rho }$ and $\Gamma _{\sigma }$, which describe Fermi-liquid renormalizations in the charge- and spin-density channels, respectively.

Up to now, averaging over disorder has been performed in a very particular fashion. Namely, within the framework of the Fermi-liquid theory, disorder has been considered only in the dynamical blocks describing diffusive propagation of electron-hole pairs interrupted by rescattering induced by the $\emph{e-e}$ interaction. This separation of the $\emph{e-e}$ interaction and disorder is the main approximation of the disordered Fermi-liquid theory. 

The content of this Section is a byproduct of a more general theory of the $e$-$e$ interactions in disordered conductors developed by Finkel’stein in Refs.~\cite{AF1983,AF1984}.

\section{Beyond Fermi-liquid theory: non-linear sigma model and
$\emph{scale-dependent}$ theory of the disordered electron liquid}
\label{sec:nlsm}

\begin{figure}[h]
\begin{flushright}\begin{minipage}{1\textwidth}  \centering
        \includegraphics[width=0.8\textwidth,clip=]{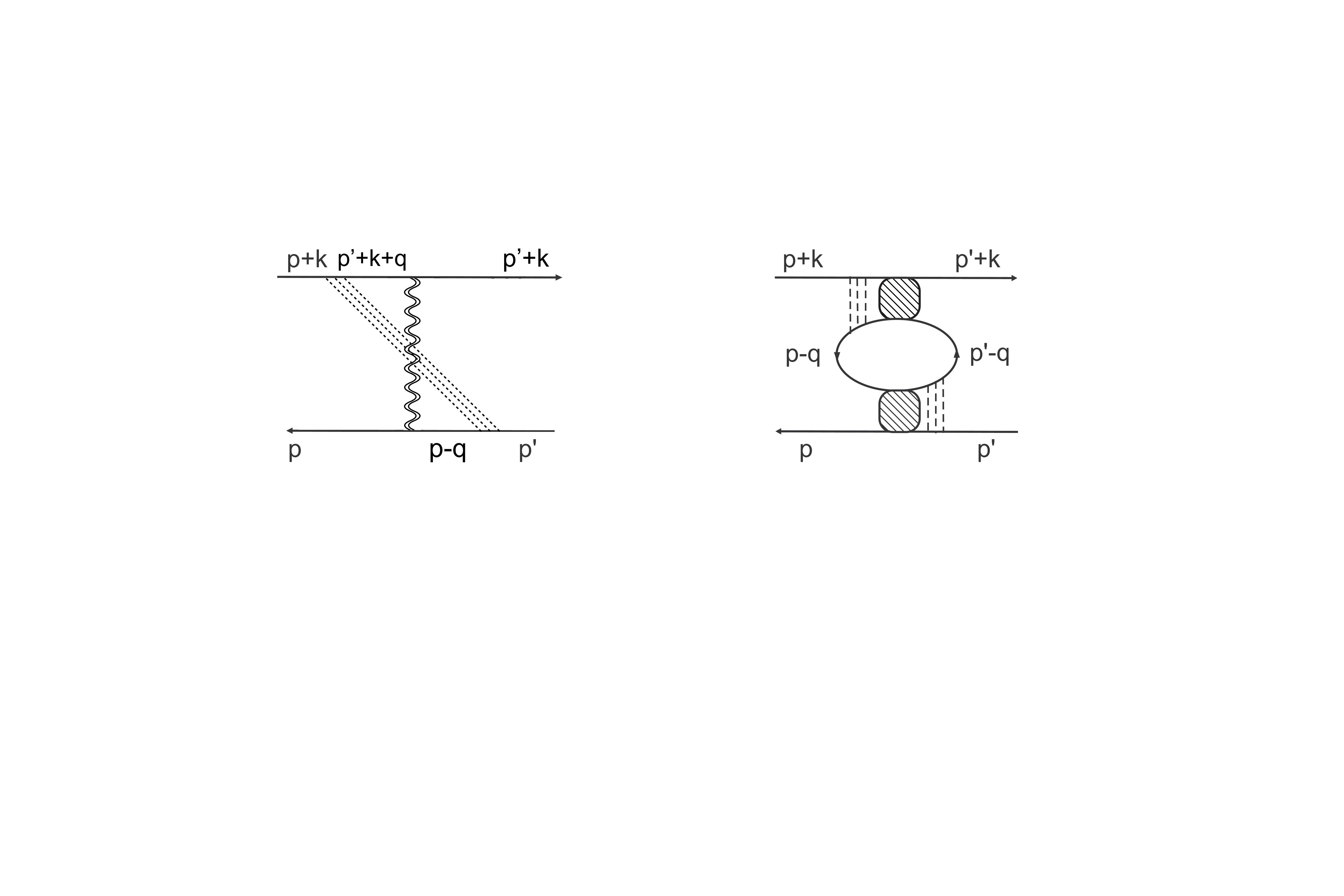}
        \caption{\small Examples of the $e$-$e$ interaction amplitudes modified by disorder.  \label{fig:examples}}
    \end{minipage}
\end{flushright}
\end{figure}

\begin{figure}[h]
\begin{flushright}\begin{minipage}{1\textwidth}  \centering
        \includegraphics[width=0.45\textwidth,clip=]{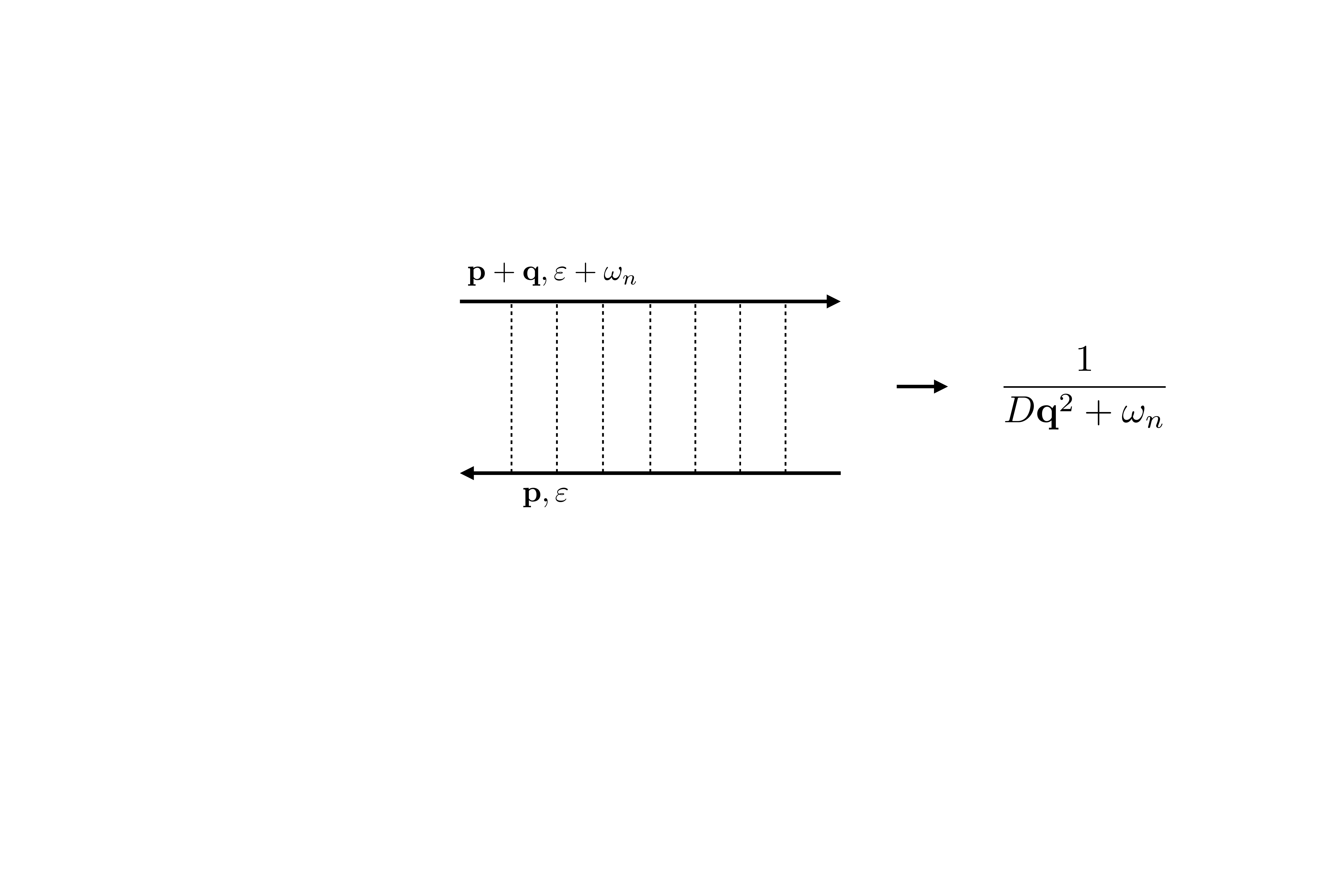}
        \caption{\small Diffuson: propagators that capture the diffusive propagation of the electron-hole pairs at large time and length scales are called "diffusons". In the figure on the left, the solid lines represents retarded and advanced disorder-dressed Green's functions. \label{fig:diffuson}}
    \end{minipage}
\end{flushright}
\end{figure}

Matrix elements determining amplitudes of the $e$-$e$ interaction are strongly modified by disorder, especially for states that are close in energy. Two examples for the modification of $e$-$e$ interaction amplitudes by impurity scattering are shown in Fig.~\ref{fig:examples}.
The amplitudes are decorated by ladders of dashed lines which result \emph{after averaging} over impurity scattering. Each of these ladders  generates a diffusion propagator $1/(Dq^2+\omega_n)$ and, consequently, is called "diffuson", see Fig.~\ref{fig:diffuson} (When frequency is on the real axis, the diffuson is equal to $1/(Dq^2-i\omega)$.]). The origin of the diffusion propagator has been described in connection with Eq.~\eqref{diffusive}. Diffusons, being singular, play a special role in modifying (renormalizing) the interaction amplitudes. The diffusons responsible for the disorder dressing in Fig.~\ref{fig:examples} are integrated either over momentum ${\bf q}$ only, as in the example shown in the left panel, or over momentum \emph{and} frequency, as in the right panel.
In the diffusive transport regime relevant momenta ${\bf q}$ lie within an interval determined by the condition $1/\tau _{el}>Dq^{2}\gtrsim T$. The temperature $T$ acts as a low-energy cut-off due to the smearing of the electronic energy distribution. Eventually, this is the reason why the integration over momenta [N.6] can result in corrections to the $e$-$e$ amplitudes that are non-analytic in temperature~\cite{AF1983,AF1984,Cast}.

\begin{figure}[h]
\begin{flushright}\begin{minipage}{1\textwidth}  \centering
        \includegraphics[width=0.45\textwidth,clip=]{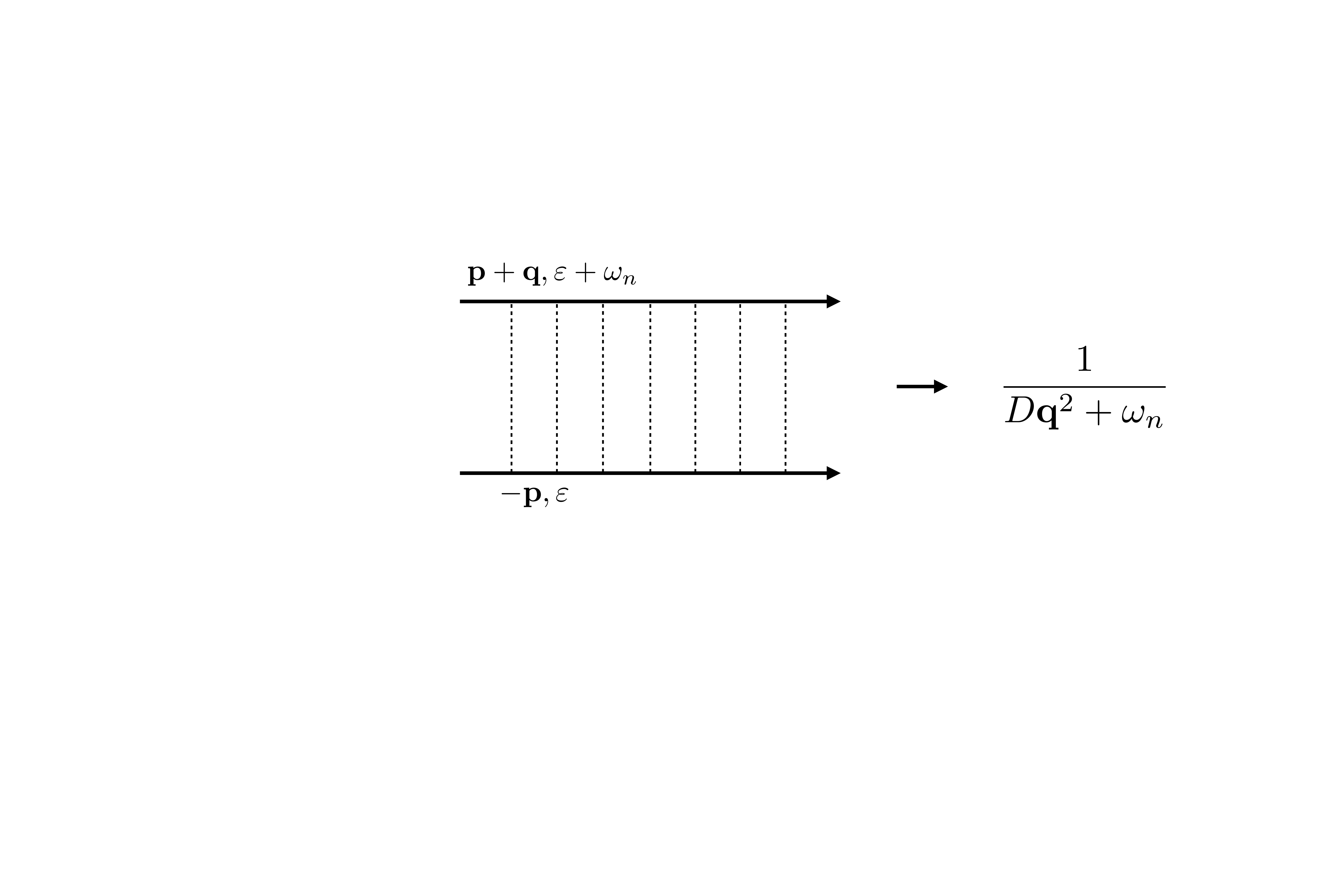}
        \caption{\small Cooperon: disorder-averaged propagator of a particle-particle pair with small total momentum. The Cooperons capture the effects of quantum interference which lead to the weak-localization corrections.\label{fig:cooperon}}
    \end{minipage}
\end{flushright}
\end{figure}

Besides the diffuson there is another mode, called the Cooperon, which also generates non-analytic corrections of \emph{various} kinds \cite{AALR,GLK}. This mode is a  "cousin" of the diffuson, and it has the same diffusion propagator. The Cooperon represents a particle-particle pair with small total momentum propagating in a disordered environment, see Fig.~\ref{fig:cooperon}. It is called the Cooperon, because of its obvious relevance for the Cooper channel in which the superconducting instability develops. 

\begin{figure}[h]
\begin{flushright}\begin{minipage}{1\textwidth}  \centering
        \includegraphics[width=0.8\textwidth,clip=]{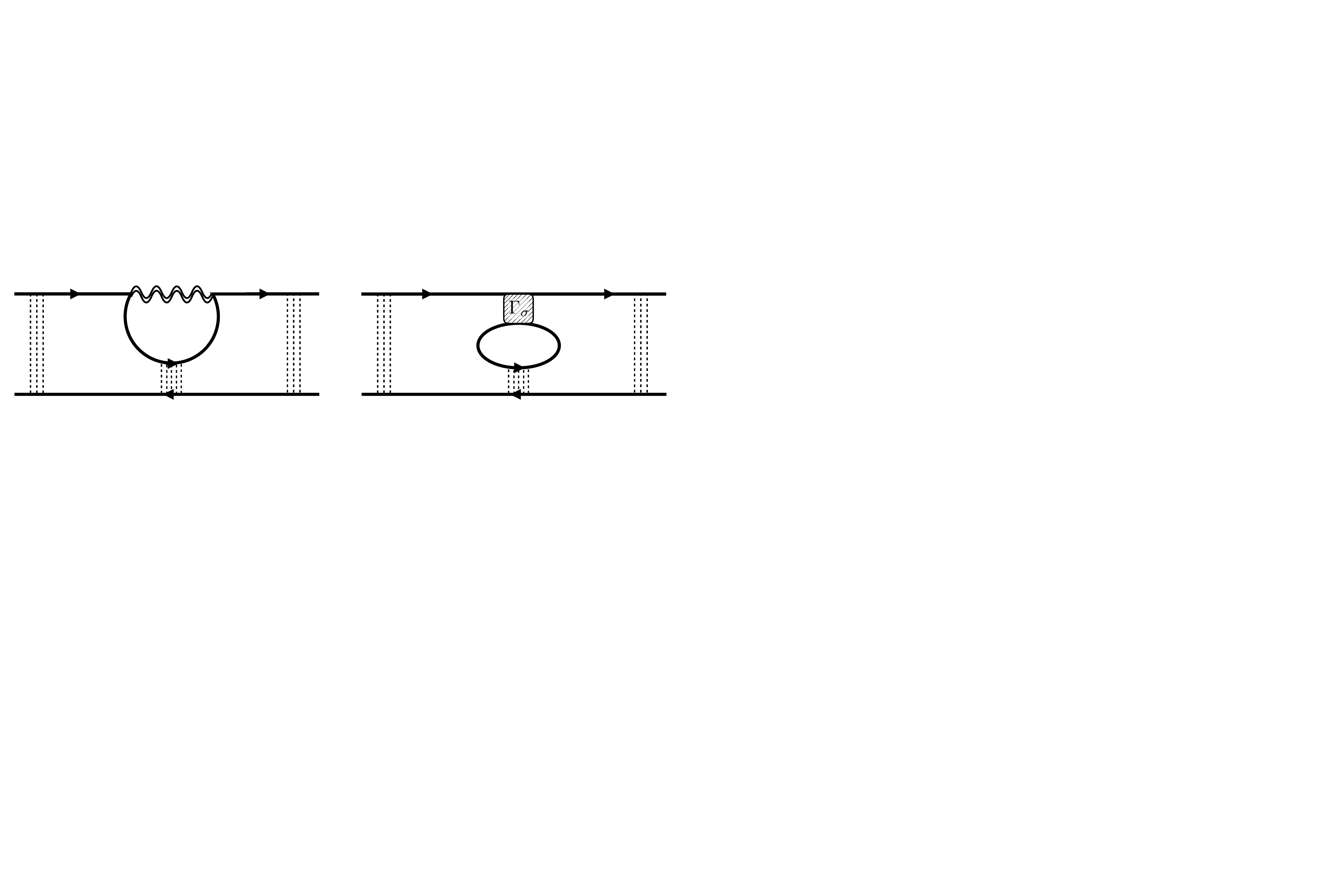}
        \caption{\small Examples of diagrams contributing to the corrections to the diffusion constant $D$ due to the interplay of disorder and interactions.} \label{fig:AALee}
    \end{minipage}
\end{flushright}
\end{figure}

The electric conductivity~\cite{AALee,AAbook} and, correspondingly, the diffusion constant $D$ also acquire corrections that are non-analytic in temperature~\cite{AF1983}, due to the combined action of the $e\text{-}e$ interaction and disorder in the diffusive regime [N.7]. The two diagrams displayed in Fig.~\ref{fig:AALee}, among others, contribute to the corrections to the diffusion coefficient $D$. 

In $d=2$, corrections \emph{both} to the diffusion coefficient (electric conductivity) and the interaction amplitudes are
logarithmically divergent in temperature, i.e., proportional to $\ln 1/T\tau_{el}$. Logarithmic corrections to the conductivity caused by the $e$-$e$ interaction were first obtained by Altshuler, Aronov $\&$ Lee~\cite{AALee,AAbook} and, conversely,  to the interaction amplitudes themselves by Finkel'stein~\cite{AF1983}. In higher dimensions, the problem becomes logarithmic near the metal-insulator transition as has been discussed in Refs.~\cite{AF1983a,AF1984MIT,Cast,CastKLee}. The divergent corrections signal the breakdown of the Fermi-liquid theory in the na\"ive form presented above.
There is a need for re-summation of the divergent terms but, most importantly, there is a need to clarify the general structure of the theory.

\emph{Objectives for the renormalization group (RG) analysis:} At low temperatures, the corrections to the resistivity and the corrections to the $e$-$e$ interaction amplitudes $\Gamma_\rho$ and $\Gamma_\sigma$ are mutually coupled. The corrections to the resistivity depend on the interaction amplitudes. The corrections to the interaction amplitudes, in turn, are a function of the strength of disorder, which can be measured in terms of the resistivity. The corrections are the more significant the larger the resistance is. The RG is able to account for both effects in a consistent way. In the RG scheme, one can obtain coupled (flow) equations for the resistivity and for the interaction amplitudes. Reviews of the RG-theory in disordered conductors can be found in Refs.~\cite{AF1990,BelitzK,DiCastro} and also in Ref.~\cite{Cast}. For a review of recent advances see Refs.~\cite{Burmistrov2017,BurmistrovKh2019}.

The RG-analysis for disordered conducting fermions can most efficiently be performed using the technique of the matrix NLSM \cite{Wegner}-\cite{ELK}. This technique was originally introduced for non-interacting systems by Wegner \cite{Wegner}, who made use of replicated bosonic variables to derive the NLSM. In an important further development, Efetov, Larkin, and Khmelnitskii \cite{ELK} employed fermionic (Grassmann) variables for the derivation of the NLSM. This fermionic formalism then served as a natural starting point for an extension of the theory to interacting electronic systems. In matrix terms, the disorder-averaged $N$-replica partition function of interacting electrons reads as $\langle Z_{N}\rangle =\int d\hat{Q}~e^{-S[\hat{Q}]}$, where
\begin{eqnarray}
S[Q] &=&\frac{\pi \nu}{8}\int d^{d}r~\;\text{Tr}\;[D(\nabla \hat{Q})^{2}-4z(
\hat{\varepsilon}\hat{Q})  -\frac{\pi \nu}{16}\int d^{d}r~\{\hat{Q}(\hat{\Gamma}_{\rho }^{0}+\hat{\Gamma}_{\rho })\hat{Q}+\hat{Q}\hat{\Gamma}_{\sigma }\hat{Q}+~\hat{Q}\hat{
\Gamma}_{c}\hat{Q}\}.  \label{NLSM}
\end{eqnarray}
This functional is usually referred to as Finkel'stein's NLSM~\cite{AF1983,AF1984,AF1990}. The matrix field $\hat{Q}$ lies in a manifold limited by the constraints: $\hat{Q}^{2}=1,\hat{Q}=\hat{Q}^{\dagger },$ and Tr\ $\hat{Q}=0$. These constraints make the problem non-linear and also very non-trivial. The components of the matrix $\hat{Q}$ are defined as $Q_{n_{1},n_{2}}^{ij,\alpha \beta }$, where $
n_{1},n_{2}$ are frequency indices, $i,j$ are the replica indices, and $\alpha ,\beta$ include the spin and quaternion indices. The quaternion indices are needed to incorporate both the diffusons and Cooperons within one scheme. The replica limit, $N\rightarrow 0$, should be taken in order to perform the averaging over different realizations of disorder [N.8]. The frequency matrix is $(\hat{\varepsilon})_{n,m}^{ij,\alpha\beta}=\epsilon
_{n}\delta _{nm}\delta _{ij}\delta _{\alpha \beta }$, where $\varepsilon_{n}=(2n+1)\pi T$ are fermionic Matsubara frequencies. All the interaction terms are restricted by the energy and momentum conservation laws [N.9]. The notation for the interaction terms is symbolical. For example, we omitted the Pauli matrices acting in the spaces of spin and quaternion degrees of freedoms \cite{AF1984, AF1990}. The additional  term $\hat{\Gamma}_{c}$ describes the $\emph{e-e}$ interaction in the Cooper channel. The interaction amplitudes $\Gamma _{\alpha =\rho ,\sigma ,c}$ are
dimensionless, but elements of the forms $(\hat{Q}\hat{\Gamma}_{a}\hat{Q})$ contain a factor of $2\pi T$ in the imaginary time technique. All further discussions in this and the following sections, with the exception of Sec.~\ref{sec:Kappa}, will be based on the replica NLSM in the Matsubara frequency representation.

Besides the diffusion coefficient and the interaction amplitudes, Finkel'stein's NLSM contains the parameter $z$, which accompanies the frequency matrix in the action $S[Q]$. This parameter equals $1$ in free electron systems, and remains equal to $1$ in the Fermi-liquid description of the disordered electron liquid. However, it acquires renormalization corrections when the theory is extended beyond the limits of the disordered Fermi-liquid theory deep in the diffusive regime. The parameter enters the diffuson and Cooperon propagators as a frequency renormalization. During the course of the RG-procedure, the frequency renormalization is crucial for maintaining compatibility with the conservation laws for charge and spin. As will be discussed Sec.~\ref{sec:scaling}, the parameter $z$ plays a central role in the critical region of the metal-insulator transition, since it determines the relative scaling of the frequency and the length scale \cite{AF1983a}. 

The  saddle-point value of the matrix $\hat{Q},$ which is usually denoted as ${\hat{\Lambda}}$, is fixed by the frequency term $(\hat{\varepsilon}
\hat{Q})$ in the above action; $\Lambda _{n,m}^{ij,\alpha \beta }=\mbox{sign}(n) \delta _{nm}\delta _{ij}\delta _{\alpha \beta }$. It is clear that for small $\varepsilon _{n}$ the fluctuating matrix $\hat{Q}$ is only weakly fixed to its equilibrium position $\hat{\Lambda}$ and, correspondingly, fluctuations are strong [N.10]. The fluctuations of the $\hat{Q}$-field are diffusons and Cooperons, which can be obtained by expanding the first two terms in $S[Q]$ up to quadratic order in $\delta \hat{Q}$=($\hat{Q}-\hat{\Lambda})$ and take the form $1/(Dk^{2}+z\omega _{n})$. 

The functional $S[Q]$ describes disordered interacting electrons with energies smaller than $1/\tau _{el}$. The Fermi-liquid description of the disordered electron liquid presented in Sec.~\ref{sec:FL} is obtained from a quadratic expansion of the action $S[Q]$ in
deviations of $\hat{Q}$ from $\hat{\Lambda}$ [N.11]. The coefficients in the action $S[Q]$ incorporate the Fermi-liquid renormalizations. The disorder-averaged interaction amplitudes, the diffusion coefficient $D$ as well as the parameter $z$ become scale-dependent at energies $\lesssim 1/\tau _{el}$. From the RG point of view, the Fermi-liquid renormalizations are obtained by ``integrating out'' high-energy states down to the scale $\sim1/\tau_{el}$. Subsequently, in the interval between $1/\tau_{el}$ and the temperature, the system is described in terms of diffusion modes and their interaction, and renormalization corrections are determined by anharmonic terms in the action $S$. Note that the decoupling of different interaction channels occurs only on the level of the Fermi liquid, i.e., only for the quadratic form of the action $S[Q]$. During the course of the RG-procedure the density-, spin- and Cooper-channels, $\alpha=\rho,\sigma,c$, interfere. An example is provided on the left side of Fig.~\ref{fig:examples}, where the conversion of the amplitude $\Gamma _{\rho }^{0}$ into $\Gamma _{\sigma }$ caused by disorder is illustrated [in this connection, see the comments on the $\beta$-functions below Eq.~(\ref{gammaceq}) and on the influence of the Coulomb interaction on the amplitude $\gamma_2$ below Eq.~(\ref{oneloopgam1})].  

With the frequency renormalization parameter $z$ included, the action $S[Q]$ preserves its form during the course of RG-transformations. With this at hand, the analysis of density-correlation functions of conserved quantities based on $S[Q]$ includes several steps: one needs  to find (i) the RG-modified static part of the correlation function, and (ii) the renormalized triangle vertex, and then to obtain (iii) the dynamical part of the correlation function with the use of the quadratic expansion of the already renormalized action $S[Q]$. Performing these steps, one may observe that it is sufficient to substitute in all Fermi liquid expressions, starting from Eq.~(\ref{ladder}) till Eq.~(\ref{spinsuscept}), the standard combinations $(1-\Gamma _{a})$ by the factors $(z-\Gamma _{a})$ that incorporate the renormalized values of the interaction amplitudes \cite{AF1984, CastellaniSpin}: 
\begin{eqnarray}
\chi _{a}(k,\omega ) &=&\chi _{a}^{static}\ \frac{Dk^{2}}{Dk^{2}+(z-\Gamma
_{a})\;\omega _{n}} =\chi _{a}^{static}\ \frac{D_{a}k^{2}}{\omega _{n}+D_{a}k^{2}}~,\hspace{0.15cm}a =\rho ,\sigma .\label{corrfunctions}
\end{eqnarray}
Here, the diffusion coefficients are defined as $D_{a}=D/(z-\Gamma _{a})$. For $\chi_{a}(k,\omega )$ to acquire this form the following relation has to be fulfilled during the course of renormalizations:
\begin{equation}
\frac{\chi _{a}^{static}}{\chi _{a}^{0}}=\frac{\left( \gamma ^{a}\right) ^{2}
}{(z-\Gamma _{a})}~.  \label{condition}
\end{equation}
It follows from Eq.~(\ref{corrfunctions}) that the relation $\sigma/e^{2}=2\nu D$,
see Eq.~(\ref{eqn:Einsteincharge}), still holds albeit with the value of $D$ obtained from the RG procedure. The observation about equality of the charge and spin conductivities, $\frac{\sigma _{charge}}{e^2}= \frac{\sigma _{spin}}{(g_L^0\mu _{B}/2)^{2}}$, also remains valid. 

\emph{The long-range Coulomb interaction}~\cite{AF1983, AF1983a}: because of the singular behavior of the Fourier component $V_{C}(k)$ at small momenta, the effective $\emph{e-e}$ interaction of the screened Coulomb interaction is equal to
\begin{equation}
\Gamma _{\rho }^{0}=2\nu \frac{\left( \gamma ^{\rho }\right) ^{2}}{\Pi _{st}}=\frac{1}{1+F_0}.
\end{equation}
Here, the factor $\left( \gamma ^{\rho }\right) ^{2}$ originates from attaching triangular vertices $\gamma ^{\rho }$ to the ending points of the screened
Coulomb interaction. Note that all the RG-corrections cancel out for $\Gamma _{\rho }^{0}$. Furthermore, the \emph{short-range} part of the interaction amplitude, i.e., the amplitude $\Gamma _{\rho }$, and the parameter $z$ are renormalized in a concurrent way, such that the resulting interaction amplitude acting in the $\rho$-channel is equal to $z$:
\begin{equation}
\Gamma _{\rho }^{0}+~\Gamma _{\rho }=z.  \label{Finvariance}
\end{equation}
This result is of fundamental importance. It demonstrates that the theory of the disordered electron liquid with Coulomb interactions displays a high degree of internal symmetry. This symmetry was examined in detail by Pruisken and his coauthors, who established a connection with the so-called $\mathcal{F}$-invariance of the theory \cite{PruiskenInvariance}, a notion which is intimately related to gauge-invariance.  

It is useful to regroup the parameters of the functional $S_Q$ by combining $z$ together with $\nu$ \cite{AF1984}. Then, $z$ acquires the meaning of a renormalization of the diffusion modes with diffusion coefficient $D_{Q}=D/z$ 
\begin{equation}
\nu \Longrightarrow z\nu ,\hspace{0.5cm}D\Longrightarrow D_{Q}=D/z.
\label{zettransform}
\end{equation}
Then, the interaction amplitudes always appear as $\Gamma _{a}/z$ \cite{AF1984}. Combining $z$ with $\nu$ allows one to use dimensional arguments. It is natural to link $z\nu $ to the specific heat $c$~\cite{CastDiC,CastKLee}. Furthermore, the Einstein relation, the renormalized susceptibilities, as well as the diffusion coefficients all take a Fermi-liquid form albeit with renormalized coefficients. For thermodynamics one gets
\begin{eqnarray}
\chi _{a}^{static} &=&z\nu (1-\Gamma _{a}/z)(\chi _{a}^{0}/\nu );\label{thermodynamics1}\\
 c/T&=&z\nu({c_0}/{T\nu}),  \label{thermodynamics2}
\end{eqnarray}
while for transport quantities
\begin{align}
D_{a} &={D_{Q}}/{(1-\Gamma _{a}/z)},\;a=\rho ,\sigma,
\\
\sigma/e^{2} &=2(\nu z)D_{Q}=2\nu D. 
\label{direct}
\end{align}
Here, $\chi _{a}^{0}/\nu $ and $c_0/T\nu$ are constants [N.12]. 

\textit{Conclusion:} The NLSM with interactions is the effective low-energy field theory for the description of electrons subject to random disorder. In fact, it is not a model [N.13]. Independent of the assumptions used for the microscopical derivation of $S_Q$, it is a \emph{minimal functional} which is, nevertheless, comprehensive enough to incorporate all essential degrees of freedom and symmetries of the disordered electron liquid. This functional provides a compact but concise description that is fully compatible with all constraints imposed by conservation laws (see [N.14] for comments about the different universality classes). After inclusion of the parameter $z$, the symmetries encoded in the functional $S_Q$ ensure its renormalizability. Correspondingly, during the course of the RG-procedure, the resulting theory of the disordered electron liquid preserves the structure
of the Fermi-liquid theory, albeit with the renormalized coefficients.

\section{Scaling theory of the metal-insulator transition (MIT) in $\mathbf{d=2+\protect\epsilon }$; role of the parameter $z$ and the tunneling density of states}
\label{sec:scaling}

In the dimension $d=2+\epsilon$ each \emph{momentum integration} involving the diffusion propagators (see e.g., examples presented in Figs.~\ref{fig:examples} and \ref{fig:AALee}) generates the dimensionless parameter [$\hbar=1$ throughout the text]
\begin{equation}
\rho =\frac{r_{d}(\lambda)}{2\pi ^{2}/e^{2}}\varpropto \frac{e^{2}}{
 \sigma }\lambda ^{d-2}.  \label{gdefinition}
\end{equation}
Here $\rho \ $is equal to the resistance $r_{d}$ of a $d$-dimensional cube of side length $\sim 2\pi /\lambda$ measured in units of $2\pi ^{2}/e^{2}$ [N.15]; $\lambda $ is the momentum cutoff which decreases during the renormalization [N.16] (the idea to consider the resistance of a cube as a scaling parameter goes back to Thouless \cite{Thouless_A_50}). Eq.~(\ref{gdefinition}) corresponds to Ohm's law with the conductivity $\sigma$ depending on the scale $\lambda $.

It follows from the structure of the action $S[Q]$, when written with the help of Eqs.~(\ref{Finvariance}) and~(\ref{zettransform}), that the
RG-procedure can be performed in terms of the dimensionless resistance $\rho$ and the reduced interaction amplitudes
\begin{subequations}
\begin{equation}
\gamma _{2} =-\Gamma _{\sigma }/z=\Gamma _{2}/z, \label{gammasmall}
\end{equation}
(for repulsive interaction, $\gamma_2>0$) and
\begin{equation}
\gamma _{c} =\Gamma _{c}/z.
\end{equation}
\end{subequations}
With these variables, the set of the RG equations takes the general form \cite{AF1984,AF1984MIT}
\begin{subequations}
\begin{equation}
d\ln \rho /dy=-\frac{\epsilon }{2}+\rho \beta _{\rho }(\rho ;\gamma
_{2},\gamma _{c};\epsilon ),\hspace{0.5cm}  \label{eq1}
\end{equation}
\begin{equation}
d\gamma _{2}/dy=\rho \beta _{\gamma _{2}}(\rho ;\gamma _{2},\gamma
_{c};\epsilon ),\label{eq22}
\end{equation}
and also 
\end{subequations}
\begin{equation}
d\gamma _{c}/dy=-\gamma _{c}^{2}+\rho \beta _{\gamma _{c}}(\rho ;\gamma
_{2},\gamma _{c};\epsilon ) \label{gammaceq} .
\end{equation} 
Here, the form of the $\beta$-functions in all three equations anticipates the fact that as a result of the renormalizations the interaction amplitudes interfere with other channels. Note that the structure of Eq.~(\ref{gammaceq}) differs from the other two RG equations displayed above. For the discussion of this equation, which determines the superconducting transition temperature $T_c$, see Refs. \cite{AF1987, AF1988, AF1994} as well as the comment [N.17].

The parameter $z$ is described by a separate equation:
\begin{equation}
d\ln z/dy=\rho \beta _{z}(\rho ;\gamma _{2},\gamma _{c};\epsilon ).
\label{zet}
\end{equation}
The RG equations are formulated in terms of the logarithmic variable $y=\ln 1/[\max (D\lambda ^{2}/z,\omega_{n})\tau ]$. With this choice, the scattering rate $1/\tau$ and the temperature $T$ can serve as natural upper and lower cut-offs, respectively. It is important to note that all $\beta$-functions in the RG equations \eqref{eq1}, \eqref{eq22}, \eqref{gammaceq} and \eqref{zet} are $z$-independent. It remains to comment on the origin of the explicit $\epsilon$-dependence of the first term on the right hand side of Eq.~\eqref{eq1}. This term appears due to the dependence of $\rho$ on $\lambda^{\epsilon}$, compare Eq.~\eqref{gdefinition}. The specific form of the functions $\beta_{\rho,\gamma_2,\gamma_c,\gamma_z}$ depends on the universality class (i.e., the symmetry) 
of the system. In reality, the MIT in $3d$ is a difficult issue for analysis, because on its way to the MIT (e.g., while lowering temperature) the system may pass through a number of \emph{crossovers} from one universality class to another, for a discussion of this point see Refs.~\cite{AF1984spin}-\cite{NAleinerL} and \cite{CastKLee}.  

For the purpose of illustration, consider the MIT in a disordered system with a strong magnetic scattering induced by magnetic impurities. In such a system, belonging to the
so-called unitary symmetry class, only fluctuations in the $\rho$-density singlet channel are important. Fluctuations of the electron spin-density as well as all Cooperon modes are not gapless anymore, because of a strong spin scattering. The unitary class is an "ultimate" one, because there cannot be any further crossovers from this class. 
The parameter $\rho$ representing the resistance is described by the equation
\begin{equation}
d\ln \rho /dy=-\frac{\epsilon }{2}+\rho \beta _{\rho }(\rho ;\epsilon ).\label{eq:drdy}
\hspace{0.5cm}
\end{equation}
According to this equation, the flow of the resistance is determined by two competing contributions, which are displayed on the right hand side. The first contribution, $-\epsilon/2$, is geometric. In dimensions $d>2$, it leads to a decrease of the resistance as the temperature is lowered. The second contribution originates from the combined effect of disorder and interactions. For the case under discussion, see Ref.~\cite{AF1983a}, it is localizing, i.e. $\beta_\rho>0$. The competition between delocalizing and localizing tendencies results in an unstable fixed point at $\rho =\rho _{c}$, which determines the behavior of the conductivity in the critical region of the MIT. For the electric conductivity measured at external frequency $\omega \gg T,$ the renormalization is cut off by $\omega $ rather than $T$.

Recall that while $\rho$ is a constant at the point of the MIT, exactly at this point the conductivity vanishes at $T=0$ when $d>2$. As it follows from Eq.~(\ref{gdefinition}), in the vicinity of the transition
\begin{equation}
\sigma (\lambda )/e^{2}\varpropto \lambda ^{d-2}.\hspace{0.5cm}
\label{critical}
\end{equation}
In the $3d$ case, for example, on the metallic side of the transition the critical behavior develops when $\lambda \gg \sigma (T=0)/e^{2}$. At finite temperatures, the renormalization group flow in the critical regime of the MIT stops at a scale when
\begin{equation}
D(\lambda)\lambda ^{2}/z(\lambda )\sim T\quad \underset{using~Eq.(\ref{direct})}{
\Longrightarrow }\quad \lambda ^{d}/\nu \sim zT.\hspace{0.5cm}
\label{lengthenergy}
\end{equation}
The above relations are a consequence of (i) the form of the diffusion propagator $\mathcal{D}(k,\omega _{n})=1/(Dk^{2}+z\omega _{n}),$ and (ii) the definition of the RG-parameter $\rho$ which exhibits a fixed point. These relations also take into account the result that all renormalizations in the relation between $\sigma$ and $D$ cancel out, $\sigma /e^{2}=2\nu D$.

Thus, in order to find the temperature or ac-frequency dependence of $\sigma(T,\omega) $ at the MIT, one has to connect the momentum and energy scales in the critical region, 
\begin{equation}
\lambda^d \sim z\max [\omega ,T].\label{lamb} 
\end{equation}
However, $z$ itself is a parameter obeying the scaling equation, see Eq. (\ref{zet}). Therefore, one needs to know the critical behavior of the parameter $z$ at the transition, which is determined by the value of $\rho \beta _{z}$ at the critical point:
\begin{equation}
\widetilde{\zeta }=-(\rho \beta _{z})_{critical\ point}\label{zetab}.
\end{equation}
Recalling that in Eq. (\ref{zet}) the logarithmic variable $y=\ln 1/[\max (D\lambda ^{2}/z,\omega)\tau ]$, one can resolve the relation given by Eq. (\ref{lamb}) and in this way to find $\lambda^d\sim \max[\omega,T]^{1+\tilde{\zeta}}$. Correspondingly, the dynamic critical exponent $z_{en/m}$ connecting energy and momentum scales ($en/m$) is
\begin{equation}
z_{en/m}=\frac{d}{1+ \widetilde{\zeta } \label{betaz}}  
\end{equation}
at the point of the MIT [N.18]. For free electrons $z$ is not renormalized, $\widetilde{\zeta }=0$, and at zero temperature $\sigma(\omega )\propto \omega ^{1/3}$ for $d=3$ \cite{ShapiroAbrahams}. The $e$-$e$ interaction modifies this critical behavior of the conductivity through $\widetilde{\zeta }$ which results in $z_{en/m}\neq d$ for the critical exponent and 
\begin{equation}
\sigma (\omega ,T)\propto (\max [\omega
,T])^{\frac{d-2}{z_{en/m}}} 
\propto (\max [\omega
,T])^{\frac{d-2}{d}(1+\widetilde{\zeta })}.
\end{equation}
If $\omega \lesssim T$, the RG flow is cut off by the temperature, and
\begin{equation}
\sigma (T)
\propto T^{\frac{d-2}{d}(1+\widetilde{\zeta }
)}.
\end{equation}

This suggests that the dependence of the conductivity on frequency and temperature can be described by a single function
\begin{equation}
\sigma (T,\omega )_{critical}\propto T^{x}f(\hslash \omega /T),
\label{scalingf}
\end{equation}
where $f(r)\rightarrow const$ when $r\rightarrow 0$, and $f(r)\varpropto r^{x}$ when $r\rightarrow \infty ,$ with $x=\frac{d-2}{z_{e/m}}=\frac{d-2}{d}(1+\widetilde{\zeta})$ [N.19]. This is a typical behavior near a quantum phase transition for which the MIT is, perhaps, a primary example. The above discussion is quite general, and its outcome does not depend on the specific details of the $\epsilon$-expansion [N.20], or any other approximation: the frequency or temperature dependence of the conductivity in the critical region of the MIT is determined by the right-hand side of the separate Eq.~(\ref{zet}) at the fixed point of the transition \cite{AF1983a,AF1984MIT}. 

There is, however, a problematic feature related to the RG flow  for systems belonging to the so-called orthogonal universality class [N.21], see Eqs.~\eqref{onelooprho1} and \eqref{oneloopgam1} below. In this case the amplitude $\gamma _{2}$ diverges at a finite scale $T^{\ast }$, and afterwards the RG-calculation becomes uncontrolled, see Refs.~\cite{AF1984} and \cite{CastellaniSpin} and, in particular, comment [N.37] below. The peculiar point here is that the diffusion coefficient $D_{\sigma}=D/(z+\Gamma_2) \rightarrow 0$ together with the divergency of $\gamma_2$. This allows to suggest that on the metallic side of the MIT a finite region of concentrations exists where the fluctuations of the spin density are already localized, while the charge density fluctuations remain itinerant. There is a number of scenarios related to the divergence of $\gamma _{2}$, see [N.22]. 

Experimentally, the study of the critical behavior of the MIT is a difficult task [N.23]. The conductivity is usually fitted \cite{BelitzK} by 
\begin{equation}
\sigma (T)_{critical}\propto T^{x}f_t(|t/t_c-1|/T^y).
\label{scalingf1}
\end{equation}
As a tuning parameter $t$, the concentration of dopants, uniaxial stress, magnetic field or intensity of light in the case of persistent photoconductor can be used [N.24]. It is assumed that in the limit $u\rightarrow 0$, the function $f_t(u) \rightarrow const$, and $\sigma (T)\propto T^{x}$ at the transition. In the opposite limit $u\rightarrow \infty$, i.e., at lowest temperatures, $f_t(u)\propto u^{x/y}$. This implies that $\sigma \rightarrow |t/t_c-1|^{x/y}$ and, thereby, the critical exponent $\mu$ is equal to $x/y$. In this analysis, a fit with the same parameters $x$ and $y$ can be performed on  both sides of the MIT, metallic and insulating. As a byproduct of such a study, the fact that the MIT is indeed a quantum phase transition can be confirmed. 

The measurements of the AC conductivity $\sigma (\omega )$ are, unfortunately, very rare. In Refs.~\cite{Gruner98} and \cite{Gruner2000} the frequency and temperature dependences were studied simultaneously in amorphous niobium-silicon alloys (Nb:Si) with compositions near the MIT. The measurements observed a one-to-one
correspondence between the $\omega$- and $T$-dependent conductivity as is characteristic/typical for a quantum phase transition. The
critical exponent $x$ was found to be equal to $1/2$ for this system. This corresponds to $z_{en/m}=2$, i.e., $\sigma (T,\omega )_{critical}\propto T^{1/2}f(\hslash \omega/k_{b}T) $, see Eq.~(\ref{scalingf}). The same scaling behavior should hold for the whole universality class which the discussed system represents. Scaling analysis of DC and AC conductivity was also performed in the amorphous magnetically doped semiconductor $a$-GdxSi(1-x), see Ref. \cite{Helgren2006}, where $z_{en/m}=2$ was found.

Another important consequence of the general structure of the RG-equations concerns thermodynamics. From the very fact that the critical region is determined by the fixed point of  Eqs.~(\ref{eq1},b) and (\ref{gammaceq}), it follows that the only parameter, which continues to evolve when changing the scale during the course of the RG, is the parameter $z$. It follows from Eqs. (\ref{thermodynamics1}) and (\ref{thermodynamics2}) that the critical temperature dependence of thermodynamic quantities at the MIT is 
described by the same dynamical critical exponent $z_{en/m}$. For example, for the specific heat this leads to $c\propto zT\propto T^{d/z_{en/m}}$.

These considerations demonstrate that the critical behavior near the MIT can be found by a dimensional analysis once the structure of the NLSM with the interaction terms is established. The parameter $z$ is crucial for obtaining the correct form of the density and spin-density correlation functions from the theory. Knowledge of these correlation functions, in turn, is crucial for the calculation of the electric conductivity and spin conductivity via the Einstein relations.

Another quantity of interest that can be analyzed with the Q-technique is the tunneling density of states (TDOS) $\nu (\varepsilon )$. The TDOS exhibits a pronounced critical behavior near the MIT \cite{TDOS}. This quantity can be obtained by measuring the differential conductance $G_{j}(V)$ of a tunneling junction at a finite voltage bias $V$: $G_{j}(V)\varpropto \nu (\varepsilon=V)$. The TDOS is defined as
\begin{equation}
\nu (\varepsilon )=-\frac{1}{\pi }\mbox{Im}\int \mathcal{G}^{R}(\varepsilon,\mathbf{p})\frac{d\mathbf{p}}{(2\pi )^{d}}.
\end{equation}
This quantity is gauge dependent. It can therefore not take part in the RG-scheme, which is based on gauge invariant quantities only. The TDOS is, nevertheless, a physically meaningful quantity. Indeed, a measurement of the tunneling current is necessarily performed with respect to an external electrode. Gauge invariance is recovered once this electrode is included into the description. A strong suppression of the TDOS is frequently observed in tunneling spectra of disordered systems \cite{AltAr, AAbook}. This effect, known as the zero-bias anomaly, is a direct consequence of the interplay of Coulomb interactions and disorder. It is stronger than other effects of the same physical origin. 

It is useful to apply the $\hat{Q}$-matrix technique for the calculation of the TDOS. With this technique, $\nu (\varepsilon )$ can be expressed as a diagonal element of the product of two matrices, averaged with respect to the fluctuations:
\begin{equation}
\frac{\nu (\varepsilon )}{\nu}= \left\langle {\hat{\Lambda}}\hat{Q}\right\rangle
_{\varepsilon \varepsilon },  \label{TDOSQ}
\end{equation}
with a projection onto one spin and replica index [N.25]. The calculation of $\nu (\varepsilon )$ can be reduced to a \emph{Gaussian integration} with respect to the matrix field describing deviations of the matrix $\hat{Q}$ from $\hat{\Lambda}$. In this way, one obtains the non-perturbative expression \cite{AF1983} [N.26], 
\begin{equation}
\nu (\varepsilon )\propto\exp \left[-\frac{\rho }{4}\ln \frac{1}{|\varepsilon |\tau}\ln
\frac{\tau \omega _{0}^{2}}{|\varepsilon |}\right],  \label{exponentiate}
\end{equation}
a generalization of the perturbative results first obtained in Refs.~\cite{AALee,Zyuzin}.

The calculation of the critical exponents of the TDOS at the MIT at $d=2+\epsilon$ using the $\epsilon $-expansion was performed in Ref. \cite{AF1984MIT}. The presence of the log-square corrections to $\nu (\varepsilon )$ when $d=2$ leads to the fact that the $\epsilon $-expansion of the critical exponent of the TDOS starts from a constant. The reason is that at $d=2+\epsilon $, the factor $1/\epsilon $ replaces one of the two logs in the exponent of Eq. (\ref{exponentiate}) and cancels a factor $\epsilon $ coming the charge $\rho _{c}\varpropto \epsilon $. The log-square corrections discussed above are specific for the long-range Coulomb interaction. Such corrections cannot arise when the dynamically screened Coulomb interaction $V_{C}(k,\omega _{n})$ is replaced by a constant. 

Mesoscopic fluctuations of the local density of states cause strong point-to-point fluctuations of tunneling spectra, which can be measured in scanning tunneling microscopy experiments. The effects of $e$-$e$ interactions on a number of local quantities were considered in detail in a series of papers by Burmistrov and collaborators summarized in Ref. \cite{mesoBurmi}. 

\textit{Conclusion:}  The MIT in the disordered electron liquid is an example of a \emph{quantum phase transition} \cite{BKV, Sachdevbook}. The dynamic scaling exponent $z_{en/m}$ is controlled by the RG parameter $z$. This parameter not only describes the frequency renormalization in the NLSM, but also determines the scaling behavior of both the conductivity and thermodynamic quantities in the critical region of the MIT. In the critical region, the scaling dependence of the parameter $z$ is described by $\widetilde{\zeta }$. The structure of the theory is more general than the $\epsilon $-expansion which can be used for the calculation of the critical exponent $\widetilde{\zeta }$. 
The appearance of a finite critical exponent $\widetilde{\zeta }\neq 0$ is a direct consequence of the $e$-$e$ interaction, and it takes different values for systems belonging to different universality classes. The value of $\widetilde{\zeta}$ and, correspondingly, of the dynamical scaling exponent $z_{en/m}$ cannot be postulated from general principals, but has to be calculated based on Eq. (\ref{zetab}). TDOS is often erroneously considered as the effective density of states in the presence of the $\emph{e-e}$-interaction. Long-range fluctuations of the electric potential give rise to log-squared corrections to the TDOS, but do not contribute to other physical quantities [N.27].

\section{Superconductivity in amorphous films}
\label{sec:SC}

Superconductivity in disordered films has been studied for decades (A.I. Shal'nikov, 1940). It is a vast arena in which different mechanisms meet and compete with each other. Disorder can weaken superconductivity, or even suppress it completely. Traditionally, one distinguishes two distinct mechanisms for this effect \cite{LarkinAnnPhys1999, Sacepe2020}: (i) the first mechanism is related to the weakening of the pairing interaction between electrons (the so-called ``fermionic mechanism"), and (ii) the second mechanism assumes that electrons are paired, but phase fluctuations of the pairing field are strong enough to prevent the formation of the coherent state (the so-called ``bosonic mechanism"). 

Here, we limit the discussion to the fermionic mechanism in amorphous (i.e. homogeneously disordered, as opposed to granular) superconducting films [N.28] The partial or even complete suppression of superconductivity caused by this mechanism can successfully be described \cite{AF1987, AF1988, AF1994} on the basis of the scaling theory discussed in Sec.~\ref{sec:scaling} when applied to a $2d$ system. The main point here is that, generally speaking, superconductivity is a weak phenomenon ($T_{c0}/\varepsilon_F\ll 1$). This is a consequence of the smallness of the ``bare" amplitude $\gamma_c$. Under these circumstances, one may observe a noticeable effect of disorder on the Cooper channel interaction even when disorder is relatively weak and, therefore, the scaling theory is fully controlled. 

\begin{figure}
\centering
\includegraphics[width=5.5cm]{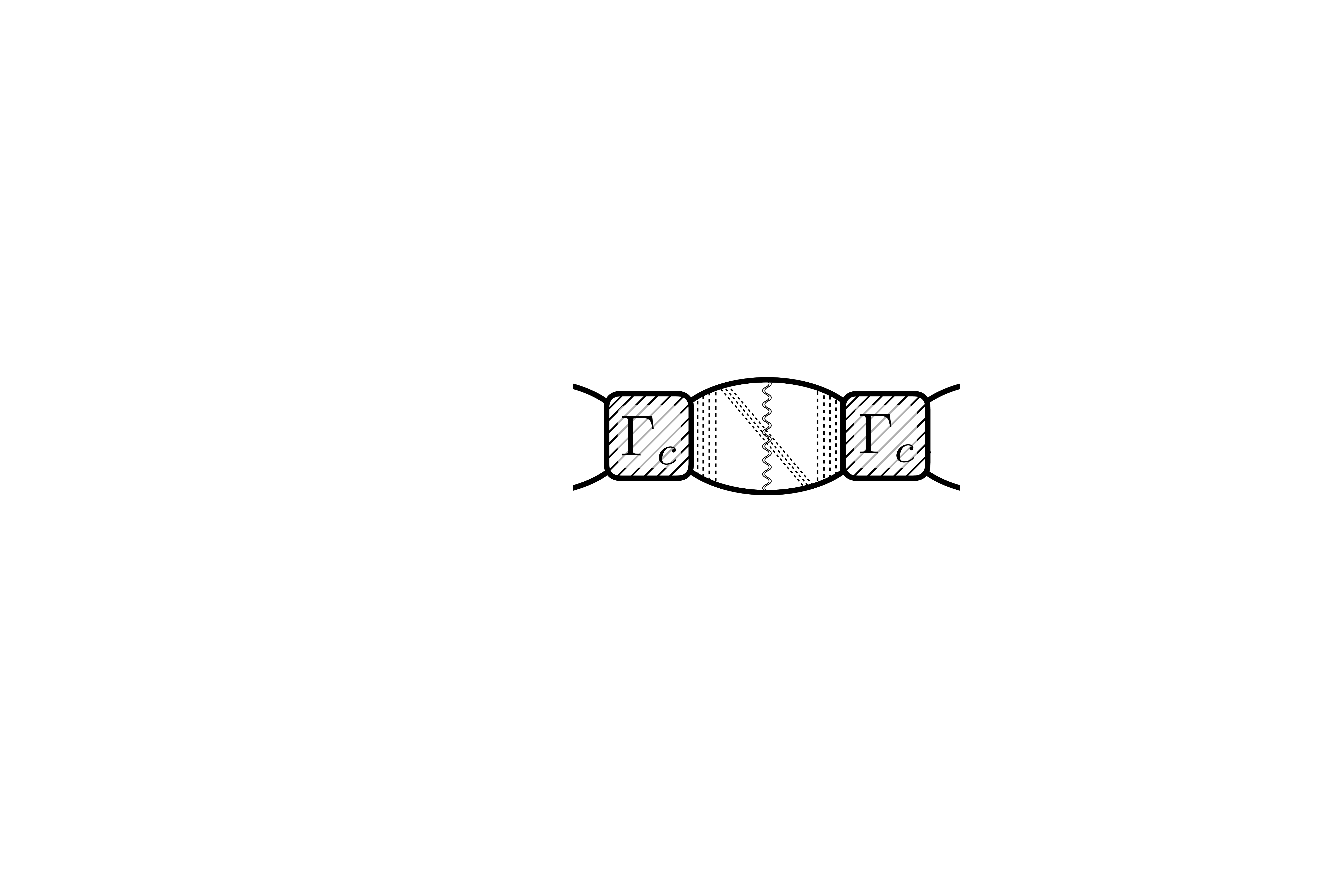}
\caption{A representative diagram for combined influence of disorder and Coulomb interactions on the Cooper channel interaction.}
\label{fig:gammacmixing}
\end{figure}

Recall that for two-dimensional films the disorder strength is conveniently characterized by the sheet resistance $R_{\square}$, which is connected to the dimensionless parameter $\rho$ as $\rho_{2d}=(e^2/2\pi^2)R_{\square}$. In the lowest order with respect to $\rho$ the function $\beta_{\gamma_c}$ in Eq.~(\ref{gammaceq}) takes the form: $\beta_{\gamma_c}(\epsilon=0)=1/2+m\gamma_2/2+\gamma_c/2$, where, in particular, the term $1/2$ accounts for the influence of the Coulomb interaction on the Cooper channel in the presence of disorder.
%\begin{equation}
%\beta_c=(1/2+m\gamma_2/2) .
%\frac{\nu (\varepsilon )}{\nu}= \left\langle {\hat{\Lambda}}\hat{Q}\right\rangle_{\varepsilon \varepsilon }, 
%\label{betac}
%\end{equation}
The factor $m$ depends on the effectiveness of the spin-orbit scattering; $m=3$ in the absence of the spin-orbit scattering, and $m=0$ when the spin-orbit scattering gaps out the triplet Cooperons and only the singlet Cooperon remains effective. In the fermionic mechanism, the charge- and spin-density channel amplitudes intervene in the Cooper channel, see Fig.~\ref{fig:gammacmixing}. Obviously, this is only possible because of the previously discussed mixing of the different channels in the presence of disorder. For the interested reader, note [N.29] comments on the relation of the fermionic mechanism to the so-called Anderson ``theorem".  

\begin{figure}
\centering
\includegraphics[width=16cm]{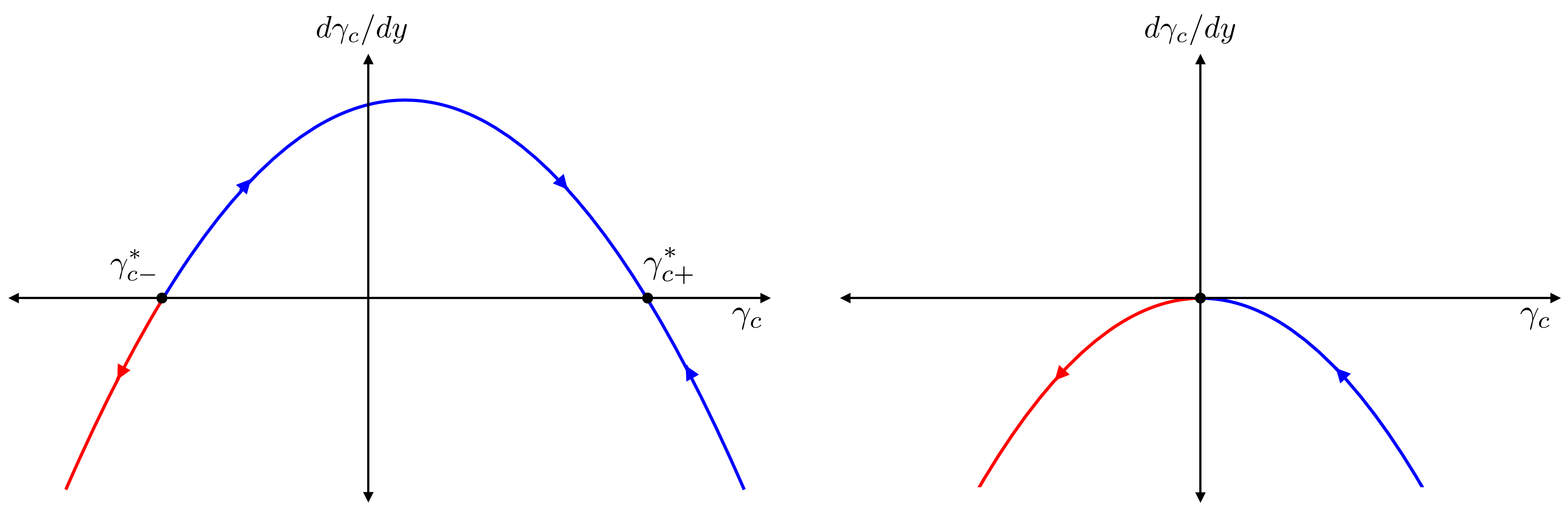}
\caption{Gell Mann Low function for disordered (left) and clean (right) superconductors. The arrows indicate the direction of the RG flow as the temperature is lowered. Systems flow towards superconductivity on the branch of the parabola shown in red, otherwise they stay normal metals.}
\label{fig:gammac}
\end{figure}

One observes that the Gell-Man Low function for the amplitude $\gamma_c$ exhibits two fixed points:
the right-hand side of Eq.~(\ref{gammaceq}) becomes equal to zero for $\gamma_{c \pm}^{*}=\pm[\rho(1/2+m\gamma_2/2)]^{1/2}+\rho/4$. 
We illustrate the RG flow in Fig. \ref{fig:gammac}. Out of the two fixed points, the positive one ($\gamma_{c+}^*$) is stable, as one can see from the directions of arrows pointing towards this point. The other one ($\gamma_{c-}^*$) is unstable. The stable fixed point corresponds to metallic behavior and is relevant not only for a repulsive interaction in the Cooper channel, as in the clean case, but also when the amplitude is attractive but not too strong. Superconductivity may only develop if $\gamma_c<\gamma_{c-}^{fp}$ holds for the ``bare" Cooper channel amplitude, due to the destructive influence of disorder. The term $\rho\beta_c$ on the right-hand side of Eq.~(\ref{gammaceq}) leads, naturally, to the suppression of the superconducting temperature $T_{\mathrm{c}}$ in homogeneously disordered films. 

The smallness of $\gamma_c$ allows us to neglect the RG-flow of $\rho$ and the interaction amplitude $\gamma_2$ during the course of the RG-analysis for the amplitude $\gamma_c$, and to consider the current values of $\rho$ and $\gamma_2$ as given. The reason for the separation of the RG-flow in the Cooper channel from the ones in the charge- and spin-density channels is that for typical $\gamma_c\lesssim \gamma_{c-}^*$, the essential range of the logarithmic variable $y$ in Eq.~(\ref{gammaceq}) is relatively small, $y\sim \rho^{-1/2}$. This has to be compared with the much larger scale $\rho^{-1}$, which is necessary for a noticeable change of $\rho$ and $\gamma_2$. As a result, $\gamma_c$ adjusts itself ``adiabatically" to the current values of $\rho$ and $\gamma_2$. 

\begin{figure}
\centering
\includegraphics[width=7.5cm]{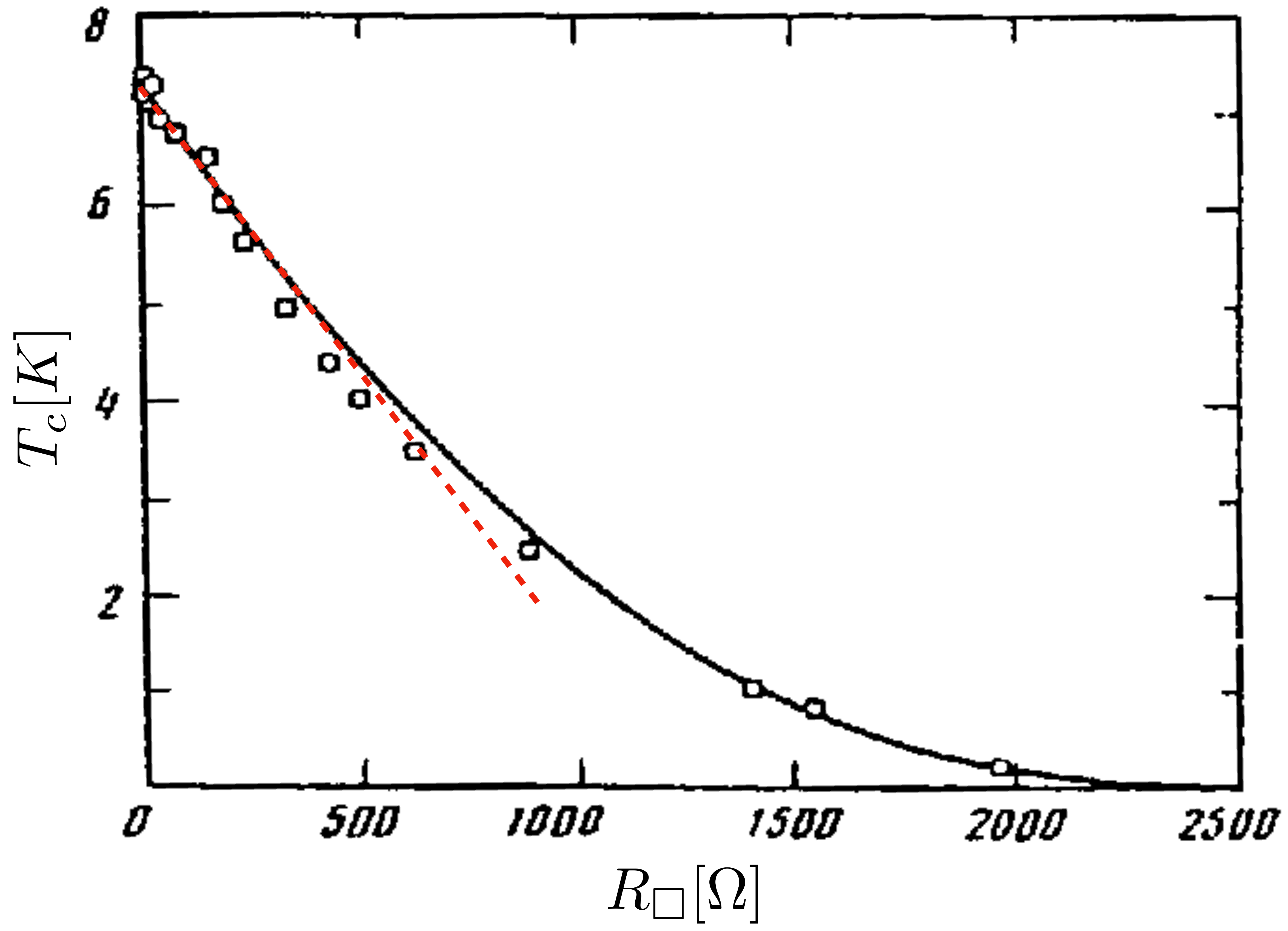}
\caption{Suppression of superconductivity in amorphous Mo$_{79}$Ge$_{21}$ \cite{GraybealBeasley1984}. The solid line is a theoretical fit obtained from the RG equation for $\gamma_c$ in Ref.~\cite{AF1987}. The red dashed line highlights the rapid initial decrease of the $T_c$-$R_\square$ curve.}
\label{fig:Tcsuppression}
\end{figure}

By fabricating a series of films with the same composition but of different thickness, one may study the effect of $\rho_{2d}$ (i.e., effective disorder) on $T_{\mathrm{c}}$, keeping aside other mechanisms. The degradation of the superconducting transition temperature $T_{\mathrm{c}}$ with disorder, see e.g., \cite{GraybealBeasley1984}, is well described by a theoretical curve obtained in \cite{AF1987}, see Fig.~\ref{fig:Tcsuppression}. The transition temperature is  found from an integration of Eq.~(\ref{gammaceq}), and is defined as the point where the amplitude $\gamma_c$ diverges. The resulting curve displays two notable features. The first one is the existence of an ending point at a relatively small resistance. For example, in amorphous samples measured in Ref.~\cite{GraybealBeasley1984} superconductivity ceases to exist when the resistance per square fulfills $R_\square\gtrsim 2.5k\Omega$. We have already discussed the reason for the suppression of superconductivity, in connection with the left of the two fixed points ($\gamma_{c-}^*$). In fact, this is an interesting example of a quantum critical point induced by disorder. The critical resistance, i.e., the resistance at the ending point of the curve, is not universal and can be noticeably smaller than the quantum resistance for superconductors $2\pi/(2e)^2\approx6.45k\Omega$. An extension of this theory to films in which other thickness-dependent
mechanisms reveal themselves \cite{XiongDynes1997} was developed in Ref. \cite{OregF1999}, see [N.30].

Another interesting feature of the disorder-tuned $T_c$ suppression is the peculiar shape of the $T_c-R_\square$ curve presented in Fig.~\ref{fig:Tcsuppression}. It describes two distinctly different dependences of $T_c$ on $R_\square$ at small and large resistances. The first one characterizes the rapid initial decrease of $T_{\mathrm{c}}$ with resistance of the film per square sheet \cite{MaekawaF1982,TakagiKuroda1982,AF1984}. (This rapid decrease is indicated in Fig.~\ref{fig:Tcsuppression} by the dashed red line.) The other one describes a very smooth decrease of $T_{\mathrm{c}}$ when the resistance is approaching the critical point. Both of the these behaviors are, in fact, encoded in a single RG-equation, Eq.~(\ref{gammaceq}). This is rather curious in view of the simplicity of this equation. The point is that we discuss $T_{\mathrm{c}}$ as a function of $\rho$. Initially, i.e., at small resistance, we deal with the parabolic Gell-Man Law function where the two fixed points only start to split and move away from each other very rapidly. Next, at larger resistance we deal with the left fixed point, which is a regular unstable point. The recalculation of the logarithmic variable $y$ at which $-\gamma_c\rightarrow\infty$ to the temperature of the transition gives a non-analytic function with a very smooth dependence of $T_{\mathrm{c}}$ on $R_\square$. 

In systems with long-range Coulomb interactions, as discussed so far, the interaction amplitude in the singlet channel is constraint by the relation stated in Eq.~\eqref{Finvariance}. When lifting this restriction, for example for systems with strong screening or cold atoms , the singlet channel interaction amplitude $\gamma_\rho$ seizes to be universal, and acquires its own RG equation. (For charged carriers, this can be realized when the conducting plane is embedded in a highly polarizable matrix which effectively suppresses the Coulomb interaction.) The full set of one-loop RG equations including short-range interactions in singlet, triplet and Cooper channes was analyzed in Refs. \cite{Burmistrov12, Burmistrov15}  [N.31]; see also the discussion of a bipartite system considered in Ref. \cite{DellAnna13}.

Finally, we would like to point out that, despite decades of studying disordered superconducting films \cite{AF1994,GoldmanMarkovic1998}, 
%the level of understanding of these systems remains far from being satisfactory. 
the subject is far from being closed .
%In homogeneously disordered films, the degradation of the superconducting temperature $T_{\mathrm{c}}$ with disorder \cite{GraybealBeasley1984} is well described by a theoretical curve obtained in \cite{Finkelstein1987}. Its notable feature is existence of an ending point, which is an example of a quantum critical point induced by disorder. However, 
In the past few years, a more complex physical picture has emerged in experiments \cite{Sacepe2020}. Namely, a gap has been found \cite{Sacepe2008} in the scanning tunneling spectroscopy measurements of the density of states of some films thought to be amorphous. This gap tends to a finite value in the vicinity of vanishing $T_{\mathrm{c}}$. This observation was attributed to inhomogeneities \cite{FeigelmanKravtsov2007,FeigelmanKravtsov2010}, and the observed gap was called a "pseudo-gap" $E_{\mathrm{g}}$. By contrast, when point-contact spectroscopy is applied \cite{Sacepe2019}, another gap, $\Delta_{\mathrm{c}}$,  is unveiled. This gap follows $T_{\mathrm{c}}$ as predicted by the fermionic mechanism, at least qualitatively, and also vanishes with $T_{\mathrm{c}}$. In \cite{Sacepe2019}, the superconducting gap $\Delta_c$ has been attributed to the phase-coherent state, while the pseudo-gap $E_g$ to preformed Cooper pairs. 
It is not clear, however, how the fermionic mechanism can be reconciled with the idea of preformed pairs. As an alternative, in Ref.~\cite{ZyuzinF2022}, the mechanism of the spin-triplet odd-frequency pairing induced by superconducting fluctuations has been considered. 

\emph{Conclusion}: The suppression of superconductivity by disorder in amorphous films is an excellent example of the applicability of the RG-analysis to disordered systems. The reason is the long-range nature of the mechanism of $T_c$-suppression. Obviously, the actual mechanism is the disorder-induced modification of the matrix elements of the Coulomb interaction for states which are close in energy. After averaging over disorder, the effect can be described as the interplay of the Coulomb interaction with the long-range Cooperon and diffuson modes. By contrast, the bosonic mechanism which is based on fluctuations of the superconducting phases of already established local superconducting order parameters depends on a different set of energy scales. Konstantin Efetov made important contributions to our current understanding of this mechanism \cite{Efetov80,Beloborodov05}.

\section{The MIT in a two-dimensional system}
\label{sec:2dMIT}
For free electrons in $2d$ the MIT (with the exception of the symplectic symmetry class) is not possible. On the contrary, the data obtained in the limit of a dilute electron liquid clearly demonstrates \cite{KP}-\cite{AKS} the existence of the MIT in a $2d$-system with a significant $\emph{e-e}$ interaction, although the mechanism of the transition has been disputed, see e.g., Ref. \cite{SKKG}.

While for studying the MIT in a $3d$ material a series of samples with a different amount of imperfections is needed, the $2d$-geometry allows to perform extensive measurements using the same device [N.32]. By applying a gate voltage, it is possible to vary broadly the density of electrons 
confined within the $2d$-channel inside the MOSFET device. When the density of electrons is high, the level of disorder is, effectively, small. Conversely, when the density is low, the electrons are at a strong level of disorder. Thus, one can pass from a metallic to an insulating behavior by varying the gate voltage, which is a big advantage of working with $2d$ systems. It has been found in Si-MOSFETs that on the metallic side of the transition the resistance $\rho (T)$ at low enough temperatures drops noticeably as the temperature is lowered further. Since (i) the resistance for a moderate strength of disorder \emph{drops} at low temperatures, while (ii) the resistance indisputably increases when disorder is strong, it is therefore \emph{unavoidable} that the MIT exists in-between (at least in $n$-(001) silicon inversion layer).

Remarkably, a weak in-plane magnetic field is sufficient to suppress the drop in the resistance in Si-MOSFETs \cite{Simonian}. This observation highlights the importance of the spin degree of freedom for the MIT, [N.33]. Another important observation is that $\rho(T)$ \emph{increases} before dropping down as the temperate is lowered. This non-monotonic behavior of $\rho(T)$ points towards the existence of a competition between different mechanisms. Therefore, any ''universal'' theory of the MIT in dilute
electron systems that emphasizes only \emph{one} aspect of the problem - most often it is a large value of the parameter $r_{s}$ - cannot describe the discussed system. Instead, material-specific properties need to be taken into account. 

\emph{Details:} 
The conduction band in an $n$-(001) silicon inversion layer has two almost degenerate valleys, $n_{v}=2$. It was suggested theoretically \cite{AFPprl} and later confirmed experimentally \cite{Kuntsevich,Mokashi} that the inter-valley scatterings is not effective in Si-MOSFET devices at the extreme conditions required for the dilute electron liquid, as long as the device still conducts electric current [N.34]. Although electrons in different valleys do not mix, the electrons of different valleys are coupled with each other by the $\emph{e-e}$ interaction amplitude $\gamma _{2}$, which is the same for all $2n_{v}$ sorts of degenerate fermions. 

The value of the critical resistance at which the transition occurred was found to be the same for devices fabricated even from different wafers although the critical density at MIT was sample-dependent \cite{KS}. This fact points toward the applicability of the RG-description of the MIT of the kind discussed above in Sec. \ref{sec:scaling}. Can the Fermi liquid be used as a starting point in a system with $r_s$ of the order of $10$? The Fermi-liquid renormalizations extracted from measurements of the Shubnikov-de Haas oscillations and the Hall coefficient in Si-MOSFETs turned out to be significant but not giant [N.35]. Thus, it makes sense to apply the RG-analysis as long as conducting electrons are degenerate and within the diffusive regime, i.e., $T\ll 1/\tau _{el}\lesssim \epsilon _{F}$ [N.36]. 

The boundary of the diffusive regime can be determined from measurements of the in-plane magnetoconductivity. The spin-splitting induced by the in-plane magnetic field suppresses the effectiveness of the spin-density fluctuations which results in a negative magnetoconductivity in the diffusive regime for $b=g_L^0\mu _{B}B/T\ll1$ \cite{LeeRmr}:
\begin{equation}
\Delta \sigma =-(e^{2}/2\pi^2)\,K_{v}C_{ee}(\gamma _{2},\rho )\,\,b^{2}.
\label{eqn:linear}
\end{equation}
In this equation, $K_{v}=n_{v}^{2}$. At not too strong disorder, $C_{ee}$ depends only on the interaction amplitude, $C_{ee}=0.091\gamma _{2}\left( \gamma_{2}+1\right)$ (Recall that $\gamma_{2}>0$ for repulsive interactions.). By contrast, at higher temperatures, when electrons are ballistic, $\Delta \sigma \varpropto (T\tau )b^{2}$. Thus, by studying the temperature dependence of the magnetoconductivity, one may chart the region where the electrons are diffusive and, furthermore, extract the temperature dependence of the amplitude $\gamma _{2}$ which appears as a result of the logarithmic corrections. 
 
Proceeding along these lines, it has been found that in Si-MOSFETs, the diffusive regime extends up to a few Kelvin in a density range around
the critical density of the MIT. The corresponding Fermi temperature is of the order of $10$ K and, therefore, electrons are already degenerate at temperatures convenient for
measurements. In addition, the $2d$ electron gas in Si-MOSFETs, which is in fact only a moderately high-mobility system, is \emph{unique} in the sense that the scattering is mostly short-range in character, so that the MIT is preceded by an extended interval of temperatures where electrons are in the \emph{diffusive regime.} On the contrary, a smooth disorder may drive a system with a high-mobility directly from the ballistic regime to the insulating phase.

In the one-loop order, the pair of coupled RG equations describing the evolution of the resistance and the amplitude $\gamma _{2}$ with temperature takes the form~\cite{AFPprl}:
\begin{subequations}
\begin{equation}
\frac{d\ln \rho }{dy}=\rho \;\left[ n_{v}+1-(4n_{v}^{2}-1)\Phi (\gamma _{2})\right] ,  \label{onelooprho1}
\end{equation}
\begin{equation}
\frac{d\gamma _{2}}{dy}=\rho \;\frac{(1+\gamma _{2})^{2}}{2}.
\label{oneloopgam1}
\end{equation}
\end{subequations}

These RG equations are valid at the lowest order in $\rho$ and, simultaneously, to all orders in the interaction amplitudes. The combination $(4n_{v}^{2}-1)$ is equal to the number of spin-valley multiplet channels. It multiplies the $n_v$-independent function $\Phi (\gamma _{2})=\frac{1+\gamma _{2}}{\gamma _{2}}\ln (1+\gamma _{2})-1$ which was first obtained in the derivation of the RG equations for $n_v=1$~\cite{AF1983}. The weak-localization corrections resulting from the Cooperon degrees of freedom enter the first RG equation through the factor $n_v$ in the square brackets. The factor $1$ in the square brackets is obtained from the long-range part of the Coulomb interaction and, therefore, universal. As for the second RG equation, Eq.~\eqref{oneloopgam1}, it is important to stress that the right hand side contains a factor of $1$ which originates from the long-range part of the Coulomb interaction as well (see the diagram presented on the left side of Fig.~\ref{fig:examples}). It is a manifestation of the mixing of different interaction channels in the presence of disorder. Due to this factor the right hand side of Eq.~\eqref{oneloopgam1} is nonvanishing even if $\gamma_2$ is initially equal to zero. As a result, the RG-flow creates a finite amplitude $\gamma_2$ in this case.  

The solution of the coupled RG equations (\ref{onelooprho1}) and (\ref{oneloopgam1}) reveals the following key features. The amplitude $\gamma_2$ increases monotonically as temperature decreases. The resistance displays insulating behavior ($d\rho /dT<0$) at high, and metallic behavior ($d\rho /dT>0$) at low temperatures. This non-monotonic behavior of the resistance is a result of a competition between charge-density diffusion modes and Cooperons, which favor localization, and the fluctuations of the spin- (and valley-) degrees of freedom, which favor delocalization. The latter stabilize the metallic state at low enough temperatures. In summary, the two RG equations act in tandem: the resistance increases the amplitude $\gamma_2$ and this increasing amplitude, in turn, changes the trend in the temperature-dependence of $\rho(T)$.

\begin{figure}[h]
\begin{flushright}\begin{minipage}{0.6\textwidth}  \centering
        \includegraphics[width=1\textwidth]{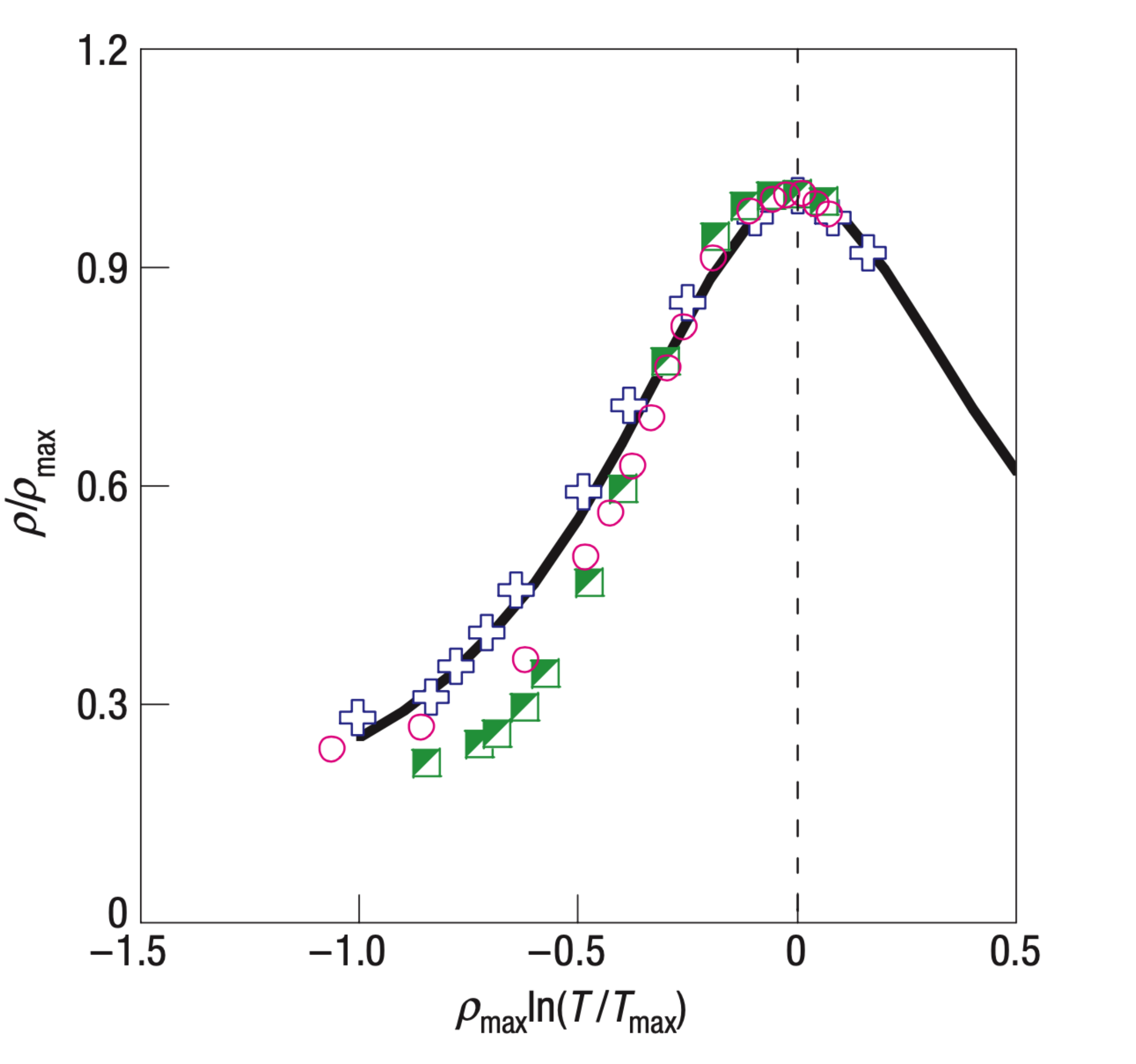}\vspace{20mm}
            \end{minipage}     \hfill
    \vspace{-20mm}\begin{minipage}[t]{0.35\textwidth} \vspace{-40mm}
     \caption{\small The RG-fitting of the resistivity in a Si-MOSFET device \cite{NatPhys}. The data corresponding to densities $n=(9.87,9.58,9.14)\times10^{10}$cm$^{-2}$
presented in Fig.~\ref{fig:MITAL} are scaled according to the universal solution given by Eq.~(\ref{universal}); no adjustable parameters have been used to fit the data. The solid line (in black) is the universal solution $R(\eta )$ of the pair of the RG equations (\ref{onelooprho1} and \ref{oneloopgam1})
with $n_{v}=2$.  
\label{fig:rhoth1}}
    \end{minipage}
\end{flushright}
\end{figure}

For $n_{v}=1$ the change in the slope of $\rho(T)$ occurs at $\gamma _{2}=2.08$, whereas for $n_{v}=2$ the combination $(4n_{v}^{2}-1)$ considerably increases the effectiveness of anti-localization, so that passing through the maximum in $\rho(T)$ happens at a considerably smaller value $\gamma _{2}=0.45$. It follows from the general form of Eqs. (\ref{onelooprho1}) and (\ref{oneloopgam1}), that the ratio $\rho (T)/\rho _{\max }$ 
is equal to a universal function, once the logarithmic variable $\eta _{T}$ is introduced \cite
{AF1983,AF1984,AF1990,AFPprl}:
\begin{eqnarray}
\rho (T)/\rho _{\max }=R(\eta _{T}) 
\qquad \eta _{T} =\rho _{\max }\ln (T/T_{\max }), 
\label{universal}
\end{eqnarray}
where $\rho _{\max }$ is the value of $\rho (T)$ at its maximum. The non-monotonic function $R(\eta )$ together with the fit of the resistance curves obtained after rescaling the data at various densities are presented in Fig. \ref{fig:rhoth1}. 
The drop of $\rho (T)$ by a factor of $5$ and the subsequent flattening of the curve at low $T$ are captured in the correct temperature interval. The full temperature-dependence of the resistance is controlled by its value $\rho _{\max }$ at the maximum; there are no other fitting parameters. Not too close to the MIT, for an intermediate range of resistances, the extracted values of $\gamma _{2}(T_{max})$ at which $\rho (T)$ passes through the maximum was found to 
corresponds to 0.45 as predicted by the theory for $n_{v}=2$. The experimental details can be found in Ref.~\cite{NatPhys}. A similar analysis of the data obtained in a device fabricated from another wafer was given in Ref.~\cite{AFPprl}; see also Ref. \cite{PudalovGamma}. 

Thus, the results obtained from the one-loop RG calculation are confirmed by the experimental data for a moderate disorder strength i.e., in the region of applicability of the one-loop approximation. Since (i) the one-loop approximation gives a \emph{drop} of the resistance at low temperatures, and (ii) the Anderson localization at strong disorder is indisputable, it is \emph{unavoidable} that the MIT exists in-between. Therefore, the existence of the MIT can be established even within the one-loop approximation [N.37]. The anti-localization effect of the $e\text{-}e$ interactions radically alters the original point of view that electrons in $2d$ are ``eventually'' (i.e., at $T=0$) always localized \cite{AALR}.

In spite of this success, Eqs. (\ref{onelooprho1}) and (\ref{oneloopgam1}) have a limited applicability. Obviously, the \emph{single-curve} (universal) solution $R(\eta )$ cannot
provide a description of the MIT. To approach the critical region of the MIT, the disorder has to be treated beyond the lowest order in $\rho $,
while adequately retaining the effects of the interaction. A consistent theory of the MIT was developed based on a two-loop calculation \cite{AFPsc} using the number of identical valleys as a large parameter, $n_{v}\rightarrow \infty$. The valley degrees of freedom are akin to flavors in standard field-theoretic models~\cite{wilson73}. It is natural, to introduce the amplitudes $\Theta =2n_{v}\gamma _{2}$ together with the resistance parameter $t=n_{v}\rho $. The parameter $t$ is thus the resistance per valley, $t=1/[(2\pi )^{2}\nu D]$. Both quantities $\Theta $ and $t$ remain finite in the large$\text{-}n_{v}$ limit.

\begin{figure}[tp]
\begin{flushright}\begin{minipage}{0.6\textwidth}  \centering
        \includegraphics[width=0.95\textwidth]{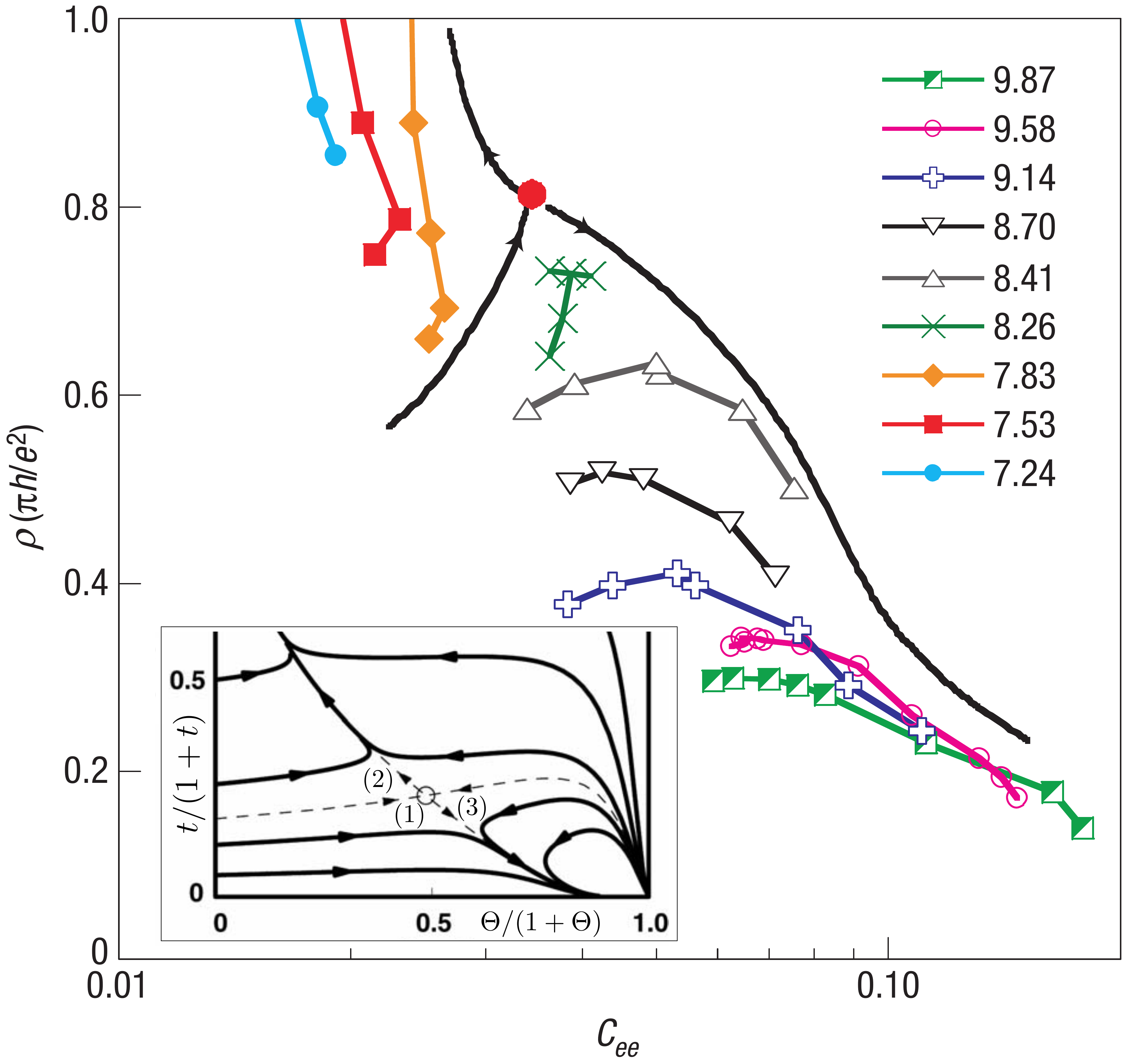}\vspace{20mm}
            \end{minipage}     \hfill
    \vspace{-20mm}
    \begin{minipage}[t]{0.35\textwidth} \vspace{-47mm}
      \caption{\small 
The resistance--interaction flow diagram of $2d$ electrons in a Si-MOSFET device \cite{NatPhys}. The location of the quantum critical point is indicated by a circle. The arrows shown on the three separatrices point in the direction of the flow for decreasing temperature. The electron densities are displayed in units of $10^{10}$ cm$^{-2}$. The coefficient $C_{ee}$ was obtained from magnetoresistance data and serves as a measure of the interaction amplitude in the spin-density channel. {\it Inset:} The inset displays the resistance-interaction ($t$-$\Theta $) flow diagram as obtained from the two-loop calculation; $\Theta =2n_{v}\gamma _{2}$ and $t=n_{v}\rho $. Area (1) is the metallic phase and area (2) is the insulating phase. Strong spin correlations are characteristic for area (3), which was not accessible in this device.}  \label{fig:MITAL} 
\end{minipage}
\end{flushright}  
\end{figure}

The resistance-interaction ($t$-$\Theta $) flow diagram is plotted as an inset in Fig.~\ref{fig:MITAL}. The arrows indicate the direction of the flow as the temperature is lowered. The quantum critical point, which corresponds to the fixed point of the equations describing the evolution of $t$ and $\Theta $, is marked by the circle. The attractive separatrices separate the metallic phase stabilized by the $e$-$e$ interaction from the insulating phase. Crossing the attractive separatrix by changing the initial values of $t$ and $\Theta $ (e.g., by changing the carrier density) leads to the MIT. The interesting region of strong spin correlations (region (3)) was not accessible in this device.

In Ref.~\cite{NatPhys} the two-parameter scaling theory has been verified experimentally. In the main panel of Fig.~\ref{fig:MITAL} the experimentally obtained flow diagram is presented. In this plot, the coefficient $C_{ee}$ effectively represents the interaction amplitude in the spin-density channel. The authors of Ref.~\cite{NatPhys} used the fact that the coefficient $C_{ee}$ reflects the strength of spin-related interactions of the diffusion modes at \emph{any} value of the resistance. Therefore, one may get broader insight into the MIT by studying the temperature dependence of the coefficient $C_{ee}$ even without knowing the exact relation connecting $C_{ee}$ with $\gamma _{2}$. This procedure has been applied for the first time in Ref.~\cite{NatPhys}, where the coefficient $C_{ee}$ has been determined by fitting $\Delta \sigma (B,T)$ to Eq.~(\ref{eqn:linear}), which allows to obtain the RG-evolution of $C_{ee}$ as a function of temperature.

The flow diagram presented in Fig.~\ref{fig:MITAL} confirms all the qualitative features of the theoretical predictions, including the quantum critical point and the non-monotonic behavior of the resistance as a function of $T$ on the metallic side of the transition. By itself, the very possibility of presenting the data as a flow diagram gives a very strong argument in favor of the two-parameter scaling theory in Si-MOSFETs.

\emph{Conclusion}: Within the region of its applicability, the RG-theory gives not only a qualitative but also a quantitative description of the experimental data in disordered electron systems, see Figs. \ref{fig:rhoth1} and \ref{fig:MITAL}. For moderate disorder, the two-parameter RG-theory therefore successfully captures the physics of the two-dimensional disordered liquid in the diffusive regime. A quantum critical point separates the metallic phase, which is stabilized by electronic interactions, from the insulating phase, where disorder prevails over the electronic interactions. The possibility of presenting the data as a flow diagram is a strong argument in favor of the applicability of the two-parameter scaling theory in a suitable systems like a Si-MOSFET. 

\section{Thermal conductivity}
\label{sec:Kappa}

The thermal conductivity $\kappa$ characterizes the heat current ${\bf j}_Q$ flowing through a material in response to a temperature gradient, ${\bf j}_Q=-\kappa\nabla T$. In contrast to the electric conductivity (see e.g., \cite{Keyes,GL}), thermal conductivity is strongly affected by inelastic processes \cite{Abrikosov59,Langer62,Lee20}. In a Fermi liquid at low enough temperatures, however, the inelastic scattering rate for particles near the Fermi energy is small, the thermal conductivity is primarily determined by disorder scattering, and the Wiedemann-Franz law (WFL) $\kappa=\mathcal{L}_0\sigma T$ with $\mathcal{L}_0=\pi^2/3e^2$ holds [N.38]. In the scale-dependent disordered electron liquid, disorder scattering and interactions become strongly intertwined. As discussed in the previous sections, their interplay leads to strong renormalizations of transport coefficients and thermodynamic quantities. The question addressed in this Section is how these renormalizations affect the thermal conductivity, and if they can lead to a violation of the WFL.

Despite its successes, the original formulation of the NLSM theory in the Matsubara formalism has certain limitations. The calculation of transport coefficients, for example, requires an analytical continuation in the complex frequency plane. Furthermore, in studying the thermal conductivity the gradient of the temperature creates a somewhat peculiar situation when the Matsubara frequencies are different in different spatial points [N.39]. The Keldysh technique~\cite{Keldish65} offers an alternative approach to the description of  interacting many-body systems. In the Keldysh technique, calculations are performed directly in real time, thereby avoiding the necessity for the analytical continuation. The range of applicability of the Keldysh approach includes equilibrium as well as non-equilibrium properties. Furthermore, the use of the Keldysh field theory approach for disordered systems is often motivated by the observation that the disorder averaging can be performed without introducing replicas~\cite{KA,Ludwig} [N.40]. The RG-analysis, including effects of the spin density fluctuations, has been reproduced in the Keldysh technique in Ref.~\cite{Schwiete14a}. The discussion of thermal conductivity in this Section will be based on the NLSM formalism in the Keldysh technique. 
 
The Keldysh partition function is equal to $Z=\int D[\psi^\dagger,\psi] \exp(iS[\psi^\dagger,\psi])$, with an action defined on the Keldysh contour $\mathcal{C}$~\cite{Keldish65,Kamenev11}. In the absence of the source terms, the action is equal to
\begin{align}
S_{k}[\psi^\dagger,\psi]=\int_\mathcal{C}dt \int_{\bf r} \left(\psi^\dagger i \partial_t\psi-k[\psi^\dagger,\psi]\right),
\end{align}
where $k=h-\mu n$ is the heat density, $h$ and $n$ are the hamiltonian density and particle density, $\mu$ is chemical potential, and $\psi$ and $\psi^\dagger$ are vectors of Grassmann fields with two spin components. The calculation of the thermal conductivity can be based on the retarded heat density correlation function $\chi_{k}(x_1,x_2)=-i\theta(t_1-t_2)\langle[\hat{k}(x_1),\hat{k}(x_2)]\rangle_T$. Here, we denoted $x=({\bf r},t)$ and introduced the heat density operator $\hat{k}=\hat{h}-\mu\hat{n}$. The angular brackets $\langle \dots \rangle_T$ stand for thermal averaging.

Introducing fields on the forward ($+$) and backward ($-$) paths of the Keldysh contour, one may define the classical ($cl$) and quantum components ($q$) of the heat density symmetrized over the two branches of the contour, $k_{cl/q}=\frac{1}{2}(k_+\pm k_-)$ \cite{Kamenev11}. After adding the source term $S_\eta=2\int_x [\eta_2(x)k_{cl}(x)+\eta_1(x)k_q(x)]$ to the action, $S=S_{k}+S_\eta$, where $\eta_1$ and $\eta_2$ are so-called gravitational potentials \cite{Luttinger64,Shastry,KarenNernst}, one can find $\chi_{k}$ by the formal differentiation of the partition function, $\chi_{k}(x_1,x_2)=\frac{i}{2}\delta^2 Z/{\delta \eta_2(x_1)\delta \eta_1(x_2)}$. The thermal conductivity $\kappa$ can be obtained from the disorder-averaged correlation function $\langle\chi_{k}(x_1,x_2)\rangle_{dis}=\chi_{k}(x_{1}-x_2)$  as \cite{Castellani87}
\begin{align}
\kappa =-\frac{1}{T}\lim_{\omega\rightarrow 0}\lim_{k\rightarrow 0}\left[\frac{\omega}{k^2} \mbox{Im}\chi_{k}(k,\omega)\right].
\label{eq:kappafromchi}
\end{align}

The disadvantage of working with the gravitational potentials  based on the action $S=S_k+S_\eta$ is that $\eta_{1,2}$ couple (among other terms) to the randomly distributed disorder potential $u_{dis}$. Fortunately, one can release the disorder term from the explicit dependence on the gravitational fields with the transformation $\psi\rightarrow \hat{\lambda}^{1/2}\psi$, ${\psi}^\dagger\rightarrow {\psi}^\dagger \hat{\lambda}^{1/2}$, where $ \hat{\lambda}=1/(1+\eta_1 \hat{\sigma}_0+\eta_2\hat{\sigma}_1)$ and $\hat{\sigma}_{0}$ and $\hat{\sigma}_1$ are Pauli matrices in Keldysh space. From here on, the NLSM can be derived along the standard lines \cite{Wegner,ELK,AF1990,KA,Ludwig,Schwiete14a}. 

\begin{figure}[h]
\begin{flushright}\begin{minipage}{1\textwidth}  \centering
        \includegraphics[width=0.50\textwidth,clip=]{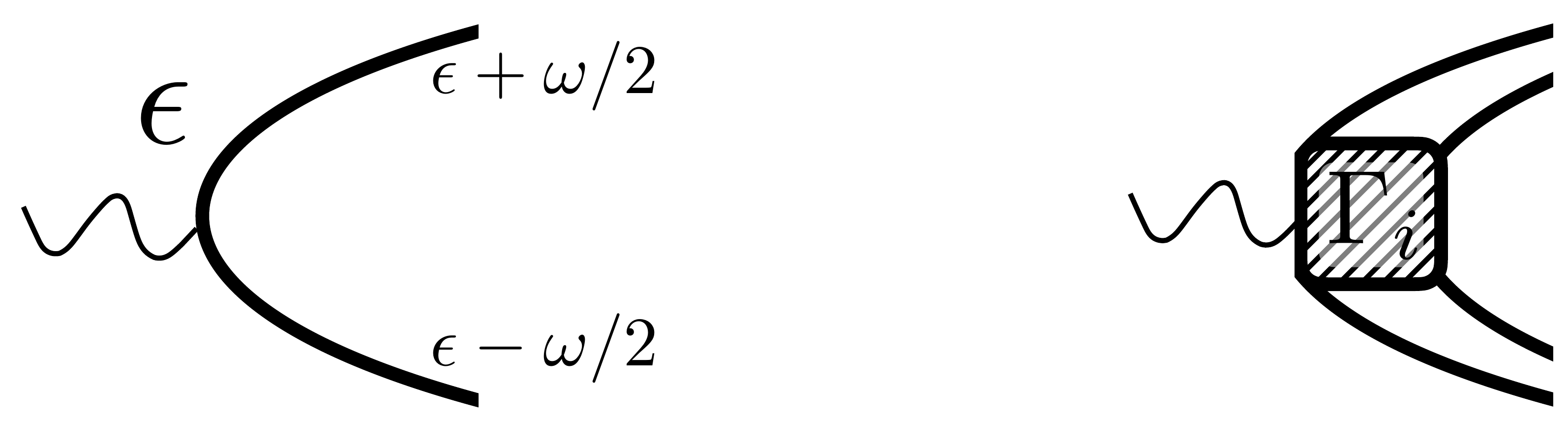}
        \caption{\small The expansion of the action of the diffusion modes $S_{dm}$ to first order in $\hat{\eta}$ gives rise to two vertices. The frequency vertex displayed on the left contains the ``center-of-mass" frequency of the two adjacent fermionic propagators. The interaction vertex on the right accounts for the presence of $\hat{\eta}$ in the interaction term.   \label{fig:heatvertices}}
    \end{minipage}
\end{flushright}
\end{figure}

\emph{Model with short-range interactions:} The diffusion modes are described by the extended NLSM [N.40] which incorporates the gravitational potentials \cite{Schwiete14b,Schwiete14c}
\begin{align}
S_{dm}=\frac{\pi\nu i}{4}\mbox{Tr}[D(\nabla\hat{Q})^2+2iz\{\hat{\epsilon},\hat{\lambda}\} \underline{\delta \hat{Q}}]-\frac{\pi\nu}{8} \int d^{d}r~\nu \{\underline{\delta\hat{Q}}\hat{\lambda}(\hat{
\Gamma}_{\rho }+\hat{\Gamma}_{\sigma})\underline{\delta\hat{Q}}\}.
\label{eq:Sd}
\end{align}
Here, $\delta\hat{Q}=\hat{Q}-\hat{Q}_{sp}$, where $\hat{Q}_{sp}=\hat{\sigma}_3$ is the (metallic) saddle point in the Keldysh NLSM. In the frequency domain, $(\hat{\lambda}_{\bf r})_{\epsilon\epsilon'}=\hat{\lambda}_{\bf r,\epsilon-\epsilon'}$, while $\hat{Q}_{\epsilon\epsilon'}$ generally depends on both frequency arguments. The trace $\mbox{Tr}$ covers all degrees of freedom including spin as well as integration over coordinates. 
For $\underline{\delta \hat{Q}}(\epsilon\epsilon')=\hat{u}_{\epsilon}\delta\hat{Q}_{\epsilon\epsilon'}\hat{u}_{\epsilon'}$ the temperature of electrons enters through the distribution function encoded in $\hat{u}$, 
\begin{align}
\hat{u}_\epsilon=\left(\begin{array}{cc} 1&F_\epsilon\\0&-1\end{array}\right),\quad F_\epsilon=\tanh\left(\frac{\epsilon}{2T}\right).
\end{align}
Similar to Eq.~\eqref{NLSM}, the structure of the interaction term in Eq.~\eqref{eq:Sd} is symbolic. At first order in $\hat{\eta}$, the gravitational potential generates two types of vertices, one associated with frequency, the other associated with the interactions. These vertices are displayed in Fig.~\ref{fig:heatvertices}.  

The static part of the heat density correlation function is related to the specific heat $c$ as $\chi_k^{static}=-cT$, while the specific heat, in turn, can be found from the average heat density as $c=\partial_T\langle \hat{k}\rangle_T$. It comprises a trivial electronic part $c^0$ and a contribution of the diffusion modes $c^{dm}$. The diffusion modes give rise to the heat density $k^{dm}=(i/2)\delta Z^{dm}/\delta \eta_2(x_1)|_{\hat{\eta}=0}$, where $Z^{dm}$ is determined by $S_{dm}$ of Eq.~\eqref{eq:Sd}. Evaluating $k^{dm}$ in the ladder approximation, the heat density can be presented in the form~\cite{Schwiete14b}
\begin{align}
k^{dm}=\frac{1}{2}\int_{{\bf q},\omega}\omega\mathcal{B}_\omega D{\bf q}^2\left[z_1\mathcal{D}_1\overline{\mathcal{D}}_1+
3z_2\mathcal{D}_2\overline{\mathcal{D}}_2-4z\mathcal{D}\overline{\mathcal{D}}\right].
\label{eq:kdm}
\end{align}
In this formula, we used several different diffuson propagators. The bare retarded diffuson $\mathcal{D}$ includes the frequency renormalization and takes the form $\mathcal{D}({\bf q},\omega)=1/(D{\bf q}^2 -iz\omega)$. In the presence of rescattering caused by the $e$-$e$ interactions, the retarded diffusion propagators in the singlet and triplet channel read as $\mathcal{D}_{1,2}=1/(D{\bf q}^2-iz_{1,2}\omega)$ with $z_1=z-\Gamma_\rho$, $z_2=z-\Gamma_\sigma$.  Advanced diffusons are written with a bar, as in $\overline{\mathcal{D}}({\bf q},\omega)=\mathcal{D}({\bf q}, -\omega)$, for example. The bosonic distribution function is $\mathcal{B}_\omega=\cot(\omega/2T)$. The form of the heat density of diffusion modes displayed in Eq.~\eqref{eq:kdm} can be understood as follows: It represents the energy weighted with the distribution function and multiplied by the spectral function of the diffusion modes. The contribution of the diffusion modes to the specific heat is obtained by differentiating the heat density with respect to temperature as $c^{dm}=\partial_T k^{dm}$. Fermions give a separate contribution to the specific heat. Eventually, an analysis of the combined fermionic and diffusion mode contributions to the specific heat leads to the conclusion that the specific heat in the disordered electron liquid can be expressed through the parameter $z$ as  $c=zc_{0}$~\cite{CastDiC,Pruisken,AF1984MIT}, where $c_{0}=2\pi^2\nu T/3$. 

For finding the thermal conductivity, it is sufficient to know the dynamical part of the heat density correlation function $\chi_{k}^{dyn}$, see Eq.~\eqref{eq:kappafromchi}. The study of $\chi^{dyn}_{k}({\bf q},\omega)$ requests for an RG-treatment in the presence of the gravitational potentials, keeping in mind their dependence on ${\bf q}$ and $\omega$. For $\chi_k^{dyn}$ it suffices to expand $\hat{\lambda}\approx 1-\hat{\eta}$ in the action. Since the RG flow for $D$, $z$, and the interaction amplitudes $\Gamma_\rho$ and $\Gamma_\sigma$ is known, it remains to determine vertex corrections by studying the renormalization of the gravitation source fields contained in $\hat{\eta}$. For the sake of simplicity, the procedure is explained for the classical component $\eta_1$. It will be preferable to work with the interaction amplitudes $\Gamma _{1}=\frac{1}{2}(\Gamma _{\rho }-\Gamma _{\sigma })$ and $\Gamma_2=-\Gamma_{\sigma}$. The NLSM \emph{extended} by the gravitational potentials has the following (symbolic) form \cite{Schwiete14b,Schwiete14c}, 
\begin{align}
S_{\zeta}=&\frac{\pi\nu i}{4} \mbox{Tr}[D(1+\hat{\zeta}_D)
(\nabla \underline{\hat{Q}})^2+2iz \{\hat{\epsilon},1+\hat{\zeta}_z \}\underline{\delta\hat{Q}}]-\frac{\pi\nu}{8} \sum_{n=1}^2\int d^d r \nu\;\{\underline{\delta\hat{Q}}(1+\hat{\zeta}_{\Gamma_n})\hat{\Gamma}_n\underline{\delta \hat{Q}}\},\label{eq:NLSMRG}
\end{align}
where $\hat{\zeta}_{X}({\bf r},\epsilon+\omega,\epsilon)=\zeta_{X}({\bf r},\omega)$ for $X\in\{D,z,\Gamma_1,\Gamma_2\}$ represent the gravitational potentials. As one can observe, in Eq.~\eqref{eq:NLSMRG} there are vertices related to frequency and interactions. Besides, the field $\zeta_D$ was introduced to anticipate the possibility that the sources $\zeta_X$ with $X\neq D$ migrate to the $D$-term during the RG procedure. The initial conditions are obtained from a comparison with Eq.~\eqref{eq:Sd}, $\zeta_z=\zeta_{\Gamma_1}=\zeta_{\Gamma_2}=-\eta_{1}$, $\zeta_D=0$, keeping in mind that only the $\eta_1$ component is used.
      
The RG-procedure would be relatively straightforward if the potentials $\hat{\zeta}_X(\epsilon,\epsilon')$ were frequency-independent. Then, the renormalization of potentials $\zeta_z$ and $\zeta_D$ could be obtained by a differentiation of the diffusion propagators as $D\partial_D\mathcal{D}$ or $z\partial_z\mathcal{D}$, while  the extraction of renormalized $\zeta_{\Gamma_n}$ could be implemented by a differentiation with respect to $\Gamma_n$. Unfortunately, $\hat{\zeta}_X(\epsilon,\epsilon')$ is a matrix in frequency space and therefore does not commute with matrices $\hat{U}$ and $\hat{\bar{U}}$ that determine the fluctuations of $\hat{Q}$ around the saddle point $\hat{\sigma}_3$. Still, the above remarks allow one to understand why the final result of the RG-analysis acquires a very compact form \cite{Schwiete14b, Schwiete14c}
\begin{align}
\Delta (X_{i_{0}}\zeta_{i_{0}})=\sum_{j=1}^4 \zeta_j \frac{\partial}{\partial X_j} (\Delta X_{i_{0}}),\label{eq:compactr}
\end{align}
where $\Delta X$ symbolizes a logarithmic correction to $X$. For the sake of notational convenience, we introduced a counting index $i$ to distinguish the RG charges $X_i\in \{D,z,\Gamma_1,\Gamma_2\}$ with $i=1...4$ as well as $\zeta_i\equiv \zeta_{X_i}$. The expression in Eqs.~\eqref{eq:compactr} resembles the so-called multiplicative RG \cite{Bogoliubov59}. The reason for this structure is that the source terms accompanying the RG charges $X_i$ are organized in a multiplicative way. It has been shown that the parameters $\zeta_{X_i}$ do not flow, $\it{provided}$ that (i) $\zeta_D=0$ holds and (ii) all remaining $\zeta_{X_i\neq D}$ are equal to each other \cite{Schwiete14b, Schwiete14c}. These criteria exactly correspond to the boundary conditions relevant for the RG-flow of the functional $S_\zeta$. Thus, the multi-parametric flow of parameters exhibits a fixed point. It is precisely this remarkable balance between the different terms in $S_\zeta$ that ensures the conservation of energy during the course of the RG-procedure.

\begin{figure}[htb]
\begin{flushright}\begin{minipage}{1\textwidth}  \centering
        \includegraphics[width=0.35\textwidth,clip=]{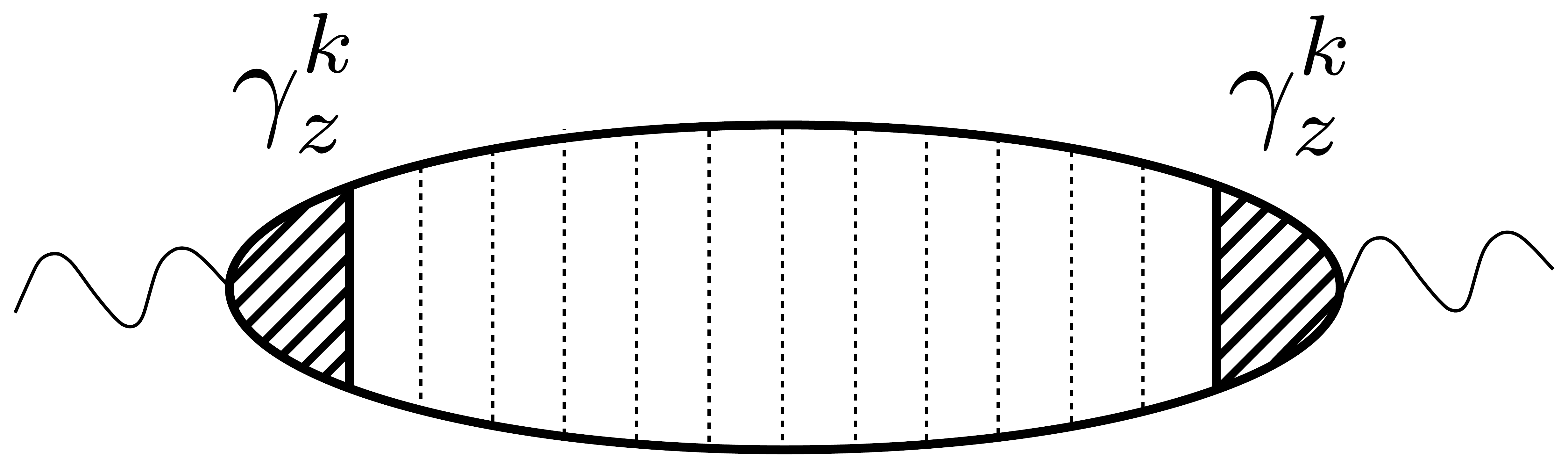}
        \caption{\small The dynamical part of the heat density-heat density correlation function. As discussed in the main text, no additional rescattering induced by the $e$-$e$ interactions occurs.        \label{fig:heatcorrelation}}
    \end{minipage}
\end{flushright}
\end{figure}

After all renormalizations, the dynamical part $\chi_{kk}^{dyn}$ is determined by the averaged product of the $\eta_{1}$- and $\eta_2$- \emph{frequency} terms that can be calculated in the ladder approximation, see Fig.~\ref{fig:heatcorrelation}. The interaction vertices are not effective at this final stage, because such contributions have already been accounted for during the RG procedure. 
Besides, unlike for the density-density or spin-density correlation functions, no rescattering of the electron-hole pairs occurs in the ladder presented in Fig.~\ref{fig:heatcorrelation} (cf. Figs. 1.3 and 1.4). The reason is that such a rescattering would induce frequency intergrals of the form $\int_{-\infty}^\infty d\epsilon\epsilon [\mathcal{F}_{\epsilon+\omega/2}-\mathcal{F}_{\epsilon-\omega/2}]=0$, unless a small particle-hole asymmetry is introduced \cite{Schwiete21}. Finally, the dynamical part of the correlation function takes the form $\chi^{dyn}(k,\omega)=-c_0 T (\gamma_z^k)^2 i\omega/(Dk^2-iz\omega)$. In this expression, the factors of $\gamma_z^k=z$ account for the renormalization of the two frequency vertices. Overall, the heat density correlation function is obtained by adding the static and dynamical parts as
\begin{align}
&\chi_{k}(k,\omega)=-c T-c_0 Tz^2\frac{i\omega}{Dk^2-iz\omega}= -cT\frac{D_Q k^2}{D_Q k^2-i\omega}.\label{eq:chidyn}
\end{align}
In these equations, the relation $c=zc_0$ for the renormalization of the specific heat has been used. With the help of Eq.~\eqref{eq:kappafromchi}, the thermal conductivity is found as $\kappa=cD_k=c_{0}D$. While the diffusion of heat is governed by the heat diffusion coefficient $D_Q=D/z$, the frequency renormalization $z$ cancels out when calculating the thermal conductivity. As an immediate consequence of this observation in combination with the RG-result for the conductivity, $\sigma=2e^2\nu D$, the WFL $\kappa/\sigma=\pi^2 T/3e^2=\mathcal{L}_0 T$ holds even in the scale dependent disordered electron liquid \cite{Castellani87,Schwiete14b}.

\emph{Coulomb interaction and violation of the WFL:} The calculation of the thermal conductivity described above is based on a model with Fermi-liquid type interactions, Eqs.~\eqref{eq:Sd} and \eqref{eq:NLSMRG}. The analysis can be extended to include long-range Coulomb interactions \cite{Schwiete16b,Schwiete16a}. The inclusion of Coulomb interactions into the RG analysis affects the results for both electric and thermal conductivities, but leaves the Wiedemann-Franz ratio $\kappa/\sigma T$ unchanged. However, the long-range character of the Coulomb interaction and the constraints imposed by gauge invariance require special care \cite{Catelani05,Schwiete16a}. For example, in the case of 2$d$ conducting electrons, while electrons are confined within a conducting plane or thin film, the electric field and its energy density extend over the whole 3$d$ space, see Sec.~III of Ref.~\cite{Schwiete16a}.

For a 2$d$ electron gas with long-range Coulomb interaction, additional logarithmic corrections to the thermal conductivity arise from energies that are smaller than temperature [N.41]. Since the sub-temperature range does not contribute to the renormalization of the electric conductivity, the WFL is violated in the presence of Coulomb interactions. The corrections captured by the conventional RG procedure result from virtual transitions. By contrast, the corrections from the sub-temperature energy range have their origin in the \emph{on-shell} scattering of electrons. The calculation of the thermal conductivity therefore demands a two-stage procedure: first, the leading logarithmic corrections originating from energies in the RG-interval $(T,1/\tau)$ can be absorbed into the scale-dependent RG charges of the \emph{extended} NLSM with gravitational potentials. Subsequently, the sub-temperature corrections can be found using the renormalized parameters of the NLSM as an input ~\cite{Schwiete16a}. The general structure of the heat density correlation function, Eq.~\eqref{eq:chidyn}, remains unaffected. However, the diffusion coefficient entering the correlation function needs to be modified as $D\rightarrow \tilde{D}=D+\delta D$, where where $D$ includes the RG-corrections and $\delta D$ originates from the sub-thermal processes.

The correction to the thermal conductivity from sub-thermal energies is found as
\begin{align}
\delta \kappa =\frac{T}{12} \log \frac{D\kappa_{scr}^2}{T}\label{eq:kappadelta}.
\end{align}
The correction $\delta \kappa$ presented in Eq. \eqref{eq:kappadelta} was obtained perturbatively by various techniques in Refs.~\cite{Raimondi04}-\cite{KarenQKE} and \cite{Schwiete16b}. The two-stage procedure described above generalized these calculations by showing \cite{Schwiete16a} that all Fermi-liquid and RG renormalizations (including the parameter $z$) drop out when calculating $\delta \kappa$. The correction $\delta \kappa$ causes a violation of the WFL. For a detailed discussion of this effect, it is useful to introduce the ``thermal resistance" $\rho_k\equiv({\mathcal{L}_0 T}/{\kappa}){e^2}/{2\pi^2}$, for which the relation $\rho_k=\rho$ holds as long as the WFL is satisfied. The thermal analog of $R(\eta_T)$ discussed in Sec.~\ref{sec:2dMIT} is 
the dimensionless function $R_k(\eta_T)=\rho_k(\eta_T)/\rho_{max}$. While the solution of the RG equations \eqref{onelooprho1} and \eqref{oneloopgam1} depends on $n_v$, the number of valleys does not affect $\delta \kappa$ \cite{Schwiete16a}. Hence, the relative weight of the WFL-violating correction is $n_v$-dependent. 
\begin{figure}
\centering
\includegraphics[width=8cm]{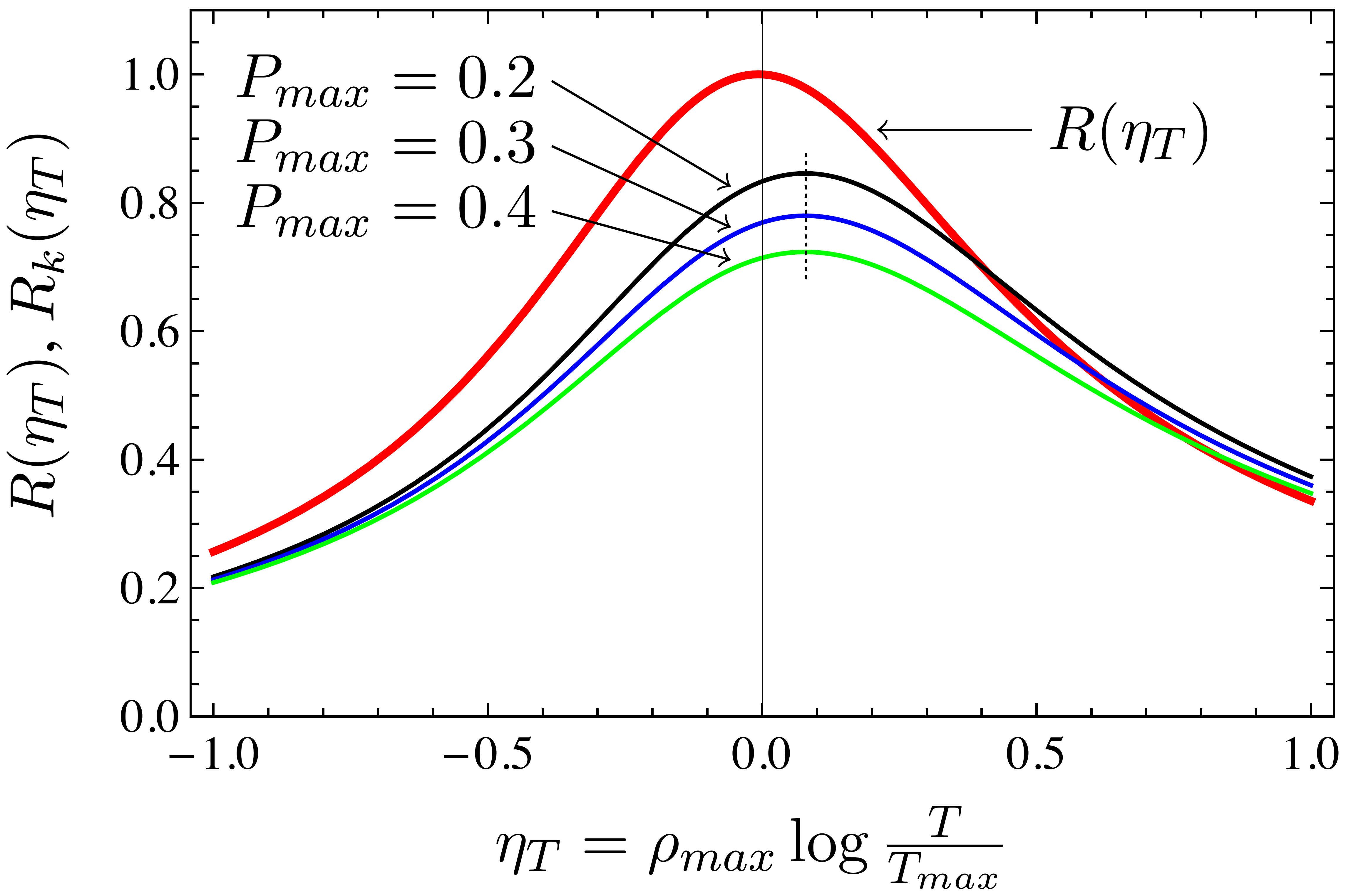}
\caption{"Thermal resistance" $R_k$ and resistance $R$ in dimensionless units for $n_v=2$. $R_k$ coincides with $R$ if the WFL holds. The parameter $P_{max}$ is a measure for the violation of the WFL law at the temperature $T_{max}$. In the discussed theory, the maxima of $R_k$ are shifted from the maximum of $R$ by a value that is independent on $P_{max}$. The high temperature-regime for which $R_k>R$ is unphysical.}
\label{fig:Rkofeta}
\end{figure}
For an arbitrary $n_v$, $R_k(\eta_T)$ can be parametrized as
\begin{align}
R_k(\eta_T)=\frac{R(\eta_T)}{1+R(\eta_T)(P_{max}-\frac{1}{2}\eta_T)},\label{eq:Rk}
\end{align}
where $P_{max}\equiv\frac{1}{2}\rho_{max}\log({D\kappa_{scr}^2}/{T_{max}})=({\rho_{max}-\rho_k(0)})/{\rho_k(0)}$ is a measure for the violation of the WFL law at the temperature $T_{max}$. Remarkably, the maximum of $R_k(\eta_T)$ is positioned at a universal value (independent of $P_{max}$), determined by the condition $R'/R^2=1/2$. For $n_v=2$, this gives $\eta^{max}_{T,k}\approx-0.0785$; examples of curves with $P_{max}=0.2$, $0.3$ and $0.4$ are shown in Fig.~\ref{fig:Rkofeta}.

Recently, it has been found that logarithmic corrections from the sub-thermal energy interval also contribute to the thermopower \cite{Jitu22}, in addition to corrections from the RG interval \cite{Fabrizio91, Jitu22}. Moreover, the sub-thermal corrections change the overall sign of the effect. The use of the NLSM formalism for the calculation of these corrections requires a generalization of the Finkel’stein model to include particle-hole asymmetry \cite{Schwiete21}.

\emph{Conclusion:} Similar to the electric conductivity, the thermal conductivity acquires a pronounced temperature dependence at low temperatures. The WFL and its validity, however, turn out to be very sensitive to the range of the interaction. For short-range interactions of the Fermi liquid-type, the WFL displays a remarkable robustness, and the low-temperature corrections to thermal conductivity closely follow those for the electric conductivity. Correspondingly, the RG captures the dominant interaction effects and the WFL is obeyed with logarithmic accuracy. In contrast, for a generic electron liquid with long-range Coulomb interactions, there exist \emph{delocalizing} contributions to the thermal conductivity that do not have any analog for the electric conductivity. They are beyond the RG approach, originate from scattering processes with very low energy transfer of the order of the temperature, and lead to a violation of the WFL.

\section{Summary}
\label{sec:Summary}
The physics of the itinerant electrons is determined by the combined effect of disorder and $e$-$e$ interactions which together generate a rich set of non-trivial phenomena. There are singular corrections to various quantities, both in transport and thermodynamics, which need to be summed consistently. The MIT is a non-trivial example of a QPT with a dynamic critical exponent which needs to be found from a special equation, rather than postulated heuristically. In two-dimensional systems, the physics is determined by the RG-flow on the disorder-interaction phase plane, which is governed by a fixed point and the associated separatrices. Crossing the separatrix implies the MIT.  In thermal transport, there exists a term violating the Wiedemann-Franz law. This violation is caused by the long-range part of the Coulomb interaction, which is effectively only time-dependent. On the methodological side, the problem is an impressive example of the effectiveness of field-theoretical methods in condensed matter physics. In particular, the originally fermionic problem is described entirely in terms of  fluctuations of the $Q$-matrix field in the effective NLSM functional, applicable to studying both normal and amorphous superconducting systems.

\section{Notes:}
\label{sec:notes}

\noindent
N.1\hspace{0.5cm}Index $k$ in $\Gamma ^{k}$ means that in the singular amplitude $\Gamma (\mathbf{k},\omega \mathbf{)}$ one first takes the limit $\omega =0$ and only afterwards the limit $k\rightarrow 0$, i.e., $\Gamma ^{k}=\Gamma (\mathbf{k}\rightarrow 0,\omega =0\mathbf{)}$.

\noindent
N.2\hspace{0.5cm}The question of the effective dimension of a disordered system has certain subtleties. Strictly speaking, two-dimensional systems are those were the motion of electrons in the third direction is quantized. In a broader sense, $2d$ systems are films where the motion of electrons is not quantized, but the diffusion modes are. In this case the \emph{effective} density of states $\nu$ incorporates the thickness of the film. Owing to the thickness, a film fabricated from a very disordered $3d$ material may have a moderate resistance. In the latter case $D=v_{F}^{2}\tau _{el}/3$. The parameter which characterizes the "effective disorder" for a film is its resistance per square, $R_{\square}$, rather than $k_fl$ or, equivalently, $\epsilon_F\tau$. The reduction to one-dimensional wires has to be performed along the same line.

\noindent
N.3\hspace{0.5cm}This implies that the zero-sound modes predicted by Landau \cite{Pitaevskii}  cease to exist in the presence of disorder.

\noindent
N.4\hspace{0.5cm}In the textbook notation, $2\widetilde{\Gamma }_{1}-\Gamma_{2}=B_{l=0}$ and $\Gamma _{2}=-C_{l=0}$; see Eqs. (18.7) and (18.9) of
chapter 2, \S 18 in Ref. \cite{Pitaevskii}.

\noindent
N.5\hspace{0.5cm} The stability of a liquid with charged particles is determined by the combination $[V_{C}(k)+\partial \mu /\partial n]>0$ (for all $k$), rather than $\partial \mu /\partial
n $ alone; the liquid is stabilized by $V_{C}(k)$, the Fourier component of the bare Coulomb interaction. 

\noindent
N.6\hspace{0.5cm}On the contrary, in the processes of rescattering presented in Figs.~\ref{fig:ladder}-\ref{fig:spindensity}, there are no integrations over the momenta of the diffusion propagators.

\noindent
N.7\hspace{0.5cm}In the ballistic region, $T>1/\tau _{el}$, non-analytic corrections due to the interplay of interaction and disorder also exist. In $2d$ they are linear in temperature, see Refs. \cite{SternDasSarma}-\cite{Zala}. The effect of a smooth long-range potential from remote ionized centers versus the influence of short-range scattering in the conducting channel was discussed in Refs. \cite{GornyiM2003} and \cite{GornyiM2004}. The experimental situation was analyzed in Ref. \cite{Clarke2008}.

\noindent
N.8\hspace{0.5cm} Besides the replica trick, a very powerful technique exists for \emph{noninteracting} systems, the supersymmetry technique pioneered by Konstantin Efetov \cite{Efetov82, Efetov97}. Its application to interacting systems, however, is challenging, and progress in this direction is so far limited \cite{Schwiete05}. An alternative to the replica trick for disordered systems \emph{with $e$-$e$ interactions} is the Keldysh formalism \cite{KA,Ludwig}. A comprehensive review of the Keldysh formalism is given in the book by A. Kamenev \cite{Kamenev11}. The whole set of results originally obtained in Ref. \cite{AF1983} was reproduced in the Keldysh technique for the first time in Ref.~\cite{Schwiete14a}.

\noindent
N.9\hspace{0.5cm}After averaging over disorder realizations, the resulting effective system is translational invariant: the scattering of the diffusion modes obeys the momentum conservation law.

\noindent
N.10\hspace{0.5cm} To get some intuition, one may look on the NLSM for non-interacting electrons as a sort of the Heisenberg functional used in the theory of magnetism. Then, the frequency term in $S[Q]$, which breaks the symmetry with respect to the sign of the frequency, is the counterpart of the external magnetic field. In this analogy, the diffusons and cooperons are the counterparts of magnons. 

\noindent
N.11\hspace{0.5cm} Systems that principally differ from the disordered Fermi-liquid are beyond the scope of this article: (i) The action describing systems such as electrons under the conditions of the quantum Hall effect, topological insulator surface states or graphene contain additional topological terms, which protect these systems from being insulators, \cite{Pruisken84}-\cite{Ostrovsky07}. Disorder may, in principle, destroy the topological protection and convert the system to an ordinary insulator. However, to this end disorder should mix opposite edges in the quantum Hall effect, or scatter electrons from one side of the topological insulator to the opposite one or mix different valleys in graphene. (ii) Another class of systems that is beyond the scope of this review are disordered non-Fermi liquid systems, i.e. systems that display non-Fermi liquid behavior in the absence of disorder \cite{Nosov20,Wu22,Patel22}. Disordered non-Fermi liquid systems may still be suited for a NLSM description, once modified to incorporate anomalous diffusion \cite{Nosov20,Wu22}.

\noindent
N.12\hspace{0.5cm}In particular, for the spin susceptibility, $\chi^0/\nu=2(g_L\mu_B/2)^2$. For the charge density, the factor $e^2$ has already been excluded from the definition of the polarization operator and, therefore, $\Pi^0/\nu=2$. For the specific heat, $c_0/(\nu T)=2\pi^2/3$.

\noindent
N.13\hspace{0.5cm}Like Landau's Fermi liquid theory is not a model. 

\noindent
N.14\hspace{0.5cm}Additional terms in the NLSM arise in the presence of magnetic impurities, an external magnetic field or spin-orbit scattering, see e.g. Ref~\cite{ELK}. In the presence of these terms, some of the diffusons become massive. At sufficiently low temperatures, when the massive diffusons may be neglected, the action of $S[Q]$ of the NLSM, Eq.~\eqref{NLSM}, can be adjusted by excluding the massive modes from the $Q$ matrix manifold. Systems with different sets of singular fluctuation propagators (i.e., when the $\hat{Q}$-fields are elements of different manifolds) belong to different universality classes. The unitary class describes systems for which only density fluctuations are important. Fluctuations of the charge- and spin-densities as well as different kind of Cooperons are relevant for the orthogonal symmetry class. In the presence of spin-orbit interactions, only the charge density mode and the singlet Cooperon remain massless, and the system belongs to the symplectic symmetry class. Besides the symmetry of the $Q$ matrix, the NLSM action for different universality classes also differs in the type and number of $e$-$e$ interaction terms. For the orthogonal symmetry class, for example, the most general form of the $e$-$e$ interaction as presented in Eq.~\eqref{NLSM} is required, while for the unitary class only the interaction in the charge density channel is relevant.  Generalizations of the $e$-$e$ interaction description in the NLSM for different symmetry classes were considered both in the Matsubara and the Keldysh techniques in Refs. \cite{delAnna} and \cite{LLF}, respectively.

\noindent
N.15\hspace{0.5cm}Compared to the standard definition of the quantum resistance, this unit contains an additional factor $\pi $. Thereby resistance $\rho$ is measured in units of $2\pi^2 \hbar/e^2$. 

\noindent
N.16\hspace{0.5cm}Decreasing $\lambda$ corresponds to enlarging blocks in the real-space renormalization procedure. 

\noindent
N.17\hspace{0.5cm} In amorphous films superconductivity can be totally suppressed by a moderate resistance per square. For a review of this question see Ref.~\cite{AF1994}.

\noindent
N.18\hspace{0.5cm}So far, one may be under the impression that parameter $z$ is inseparable from $\nu$. Eq.~(\ref{betaz}) demonstrates, however, that the situation is more delicate, because $z$ also determines the critical dynamic exponent $z_{en/m}$.

\noindent
N.19\hspace{0.5cm}In Refs. \cite{AF1983a}, \cite{AF1984MIT} and \cite{AF1990}, the combination $x=\frac{d-2}{d}(1+\widetilde{\zeta })$ was written as $\frac{d-2}{d-\varsigma }$. Correspondingly, the dynamical exponent $z_{en/m}$ has to be read as $z_{en/m}=d-\zeta$.

\noindent
N.20\hspace{0.5cm}To get an idea about the value of the critical exponent $\widetilde{\zeta\text{,}}$ consider it in the lowest order in $\epsilon $. For the case of strong magnetic scattering, the equation describing resistance in the lowest orders in $\rho $ and $\epsilon $ is
\begin{equation}
\frac{d\rho}{dy}=-\frac{\epsilon }{2}\rho+\rho^2 ,\qquad \rho _{c}=\frac{\epsilon }{2}.
\end{equation}
Furthermore,
\begin{equation}
d\ln z/dy=-\rho/2 .
\end{equation}
At $d=3$, this estimate yields $\widetilde{\zeta }=1/4$, $z_{en/m}=2.4$, and for the exponent $x\approx0.415$. The two loop calculations \cite{Pruisken,Burmistrov} at $\epsilon=1$ in the discussed case of strong magnetic scattering practically do not change the critical exponents $z_{en/m}$ and $x$; see Eq. (44) in Ref.~\cite{Burmistrov}. 

\noindent
N.21\hspace{0.5cm} This is the case when there is no mechanism limiting the fluctuations of the spin degrees of freedom, for example, when there are no impurities carrying magnetic moments which may induce spin scattering.

\noindent
N.22\hspace{0.5cm} In particular, the question of a possible transmutation of the spin diffusion modes into local ones near the MIT in $3d$ still remains unresolved. An interesting observation here is that the width of the Electron Spin Resonance line $\Delta H_{1/2}$ (i.e., the rate of relaxation) in uncompensated Si:P with $n/n_c = 1.09$ and 1.25 appears to be \emph{exactly proportional} to the spin susceptibility $\chi$ when the temperature is below 1K \cite{PSBR}. Presumably, the MIT in uncompensated Si:P differs from that in the compensated semiconductors by generating local spin modes at low enough energies in a region near the transition, for a discussion see Ref. \cite{AF1987ESR}. The discussed divergence in the spin-density channel at a finite scale may well be a driving force for the formation of local spin modes at the corresponding scale. Whether they will reveal themselves as the Kondo effect or somehow differently remains unclear, and it may depend on whether the semiconductor is compensated or not.

\noindent
N.23\hspace{0.5cm}A comprehensive discussion of experiments is beyond the scope of this article. Still a few comments seem to be appropriate here: 
 (i) The dependence of $\sigma $ on temperature in the critical region can be determined with a limited accuracy only. For the analysis of the critical behavior, the data should
be taken outside the region of perturbative corrections, $\sigma(T)-\sigma (T=0)\gtrsim \sigma (T=0)$ but, on the other hand, one should remain within the quantum transport region, $\sigma (T)<\sigma _{\min }$. (The Mott minimal conductivity $\sigma _{\min }$ is a scale separating regions where transport is dominated by classical or quantum mechanisms.) These two inequalities leave a limited window of $\sigma (T)$ appropriate for the analysis of the truly critical region of the MIT \cite{stresstuning}.

(ii) The true critical behavior may be obscured by the so-called rounding of the transition, caused by an inhomogeneous distribution of dopants. Rounding may noticeably change the critical behavior as it was demonstrated by the analysis performed in Ref. \cite{Visser}. For a discussion of an effect of rounding in Si:P see Refs. \cite{BellLab} and \cite{HvLreply}.

(iii) It appears that “canonical” Si:P is not very suitable for studying the critical behavior at the MIT. The critical region is very narrow, and several different mechanisms interfere \cite{vonL1999,HvLSiP}.

(iv) In measurements on the magnetic-field-induced MIT in GaAs and InSb semiconductors the critical behavior $\sigma (T)\sim T^{1/3}$ has been observed \cite{Pepper}. (It may be worth mentioning that for this universality class, in the lowest order in the $\epsilon $-expansion, $\widetilde{\zeta }$ is indeed equal to zero \cite{AF1984MIT}.) In a persistent photoconductor \cite{Katsumoto},  where the carrier concentration can be controlled very neatly, $\sigma (T)\sim T^{1/2}$ at the
transition, i.e., $x=1/2$; this corresponds to $\widetilde{\zeta }=1/2$. 

\noindent
N.24\hspace{0.5cm}The idea of tuning the MIT by applying a magnetic field $B$ was first put forward in Ref.~\cite{DekL}. In the case of the neutron-transmutation-doping (NTD) method, the concentration of dopants is controlled by the time of irradiation. The NTD-method allows researchers to fabricate samples with completely homogeneous doping in order to avoid rounding of the transition. Studies of the disorder driven MIT in NTD Ge:Ga semiconductors reveal a very interesting and rich picture \cite{Itoh2004}. Besides non-compensated ones, compensated samples of Ge:Ga,As with the level of compensation $K\approx0.32$ were fabricated. It was found that in non-compensated Ge:Ga at $B=0$ in a very narrow region of dopant concentrations, $(n_c-n)/n_c\lesssim 1\%$, the critical exponent is $x=0.38$. A fit with $x=0.33$ does not work well (cf. Figs. 3b and 3a in \cite{Itoh2004}).  However, when the MIT was tuned by a magnetic field, $B\neq 0$, a successful fit with the exponent $x=0.5$ can be performed in a broad region of magnetic fields. Finally, for the compensated samples in the absence of a magnetic field, $B=0$, the exponent is $x=0.33$. In all cases reasonable values of $\mu\approx 1$ were obtained.

\noindent
N.25\hspace{0.5cm}In absence of the $\emph{e-e}$ interaction, the TDOS $\nu (\varepsilon )$ is equal to $\nu$. In general, the presence of $\hat{\Lambda}$ in its explicit form (rather than as $Q$) indicates that the corresponding quantity has potential complications related to gauge-invariance, see also the next comment. 

\noindent
N.26\hspace{0.5cm}This result \cite{AF1983} has been re-derived in different ways in Refs.~\cite{Nazarov,Levitov,KA}. In particular, in Ref.~\cite{KA} the calculation of the non-perturbative TDOS was accomplished by excluding the low-momentum part of the Coulomb interaction by a gauge transformation applied directly to the matrix $\hat{Q}$. 

\noindent
N.27\hspace{0.5cm}For example, this type of correction does not influence the temperature of the superconducting transition $T_c$ (i.e., a quantity, which is very sensitive to the density of states at the Fermi energy, $\nu$). All log-squared corrections to $T_c$ are cancelled out \cite{AF1988,AF1994}. Some remnants of the time-dependent $e$-$e$ interaction reveal themselves in the thermal conductivity in $2d$; see, e.g., Refs.~\cite{KarenQKE,Schwiete16a,Schwiete16b} and also [N.41] below. 

\noindent
N.28\hspace{0.5cm}Granular superconductors consist of superconducting particles embedded in an insulating matrix. The confinement of electrons within the granules gives rise to mesoscopic energy scales that are absent in amorphous superconductors, such as the mean level spacing, charging energy and Thouless energy of individual granules, as well as their individual superconducting order parameter. Depending on the relation between these energy scales and also on the tunneling coupling between the granules, granular superconductors can display a rich behavior with superconducting, metallic and insulating phases. At strong inter-granular coupling, the granular system may resemble amorphous materials. The competition between the effect of the Coulomb blockade and the Josephson coupling between the granules has been studied by K.~B.~Efetov far ago \cite{Efetov80}. The suppression of superconductivity due to random tunneling and Coulomb effects in granular metals was addressed in Ref.~\cite{Beloborodov05}. For a review of granular electronic systems we refer to Ref.~\cite{Beloborodov07}. 

\noindent
N.29\hspace{0.5cm}According to the Anderson ``theorem" \cite{Andersontheorem,AG1959}, disorder does not influence the temperature of the superconducting transition. Obviously, the fermionic mechanism is in a stark contradiction with this theorem. For the purpose of clarity, let us emphasize that the Anderson ``theorem" is valid (and was applied by its authors) only within the scope of the BCS-Hamiltonian. In our terminology, the mixing of different channels was fully ignored in these early works. Therefore, the Coulomb interaction as well as the $\gamma_2$ amplitude could not enter the Cooper channel.

\noindent
N.30\hspace{0.5cm}The presence of non-universal mechanisms make an accurate description of the dependence of $T_{\mathrm{c}}$ on $R_{\square}$ impossible. These mechanisms can be the dependence of the electron-phonon interaction on the thickness of the films when they are too narrow, or the quantization of the density of states. However, these (unidentified) mechanisms are of a short-range character, unlike the long-range effect of the suppression of the superconductivity induced by the interplay of the Coulomb interaction and disorder. Therefore, for a wide stripe (when the width is of the order of $10^2$ nm) the non-universal effects are the same as in $2d$ films. 
Owing to this fact, the transition temperatures of the $2d$ films of different thicknesses can be used as input parameters when studying the $T_{\mathrm{c}}$-suppression for wide stripes. With this input, the $T_{\mathrm{c}}$-suppression by $R_{\square}$ 
can successfully be obtained for each of the stripes of different widths. This fact demonstrates the long-range origin of the mechanism of the $T_{\mathrm{c}}$-suppression by disorder as well as the universality of the fermionic mechanism.

\noindent
N.31\hspace{0.5cm}Specifically, Refs.~\cite{Burmistrov12, Burmistrov15} discuss a mechanism for an \emph{enhancement} of superconductivity in disordered systems with short-range interactions. For this case, the equation for the flow of the relevant amplitude $\gamma$ (which is a certain combination of the singlet, triplet and Cooper channel amplitudes), reads as $d\gamma/dy=2\rho\gamma-2\gamma^2/3$ \cite{Burmistrov12, Burmistrov15}. Unlike for the case of the Coulomb interaction discussed above, this equation needs to be studied in parallel with the flow equation for $\rho$, which for small interaction amplitudes $\gamma_i\ll 1$ is governed by the weak localization contribution, $d\rho/dy=\rho^2$. Based on the two flow equations for $\gamma$ and $\rho$, the authors of Ref.~\cite{Burmistrov12, Burmistrov15} found that if the condition $\rho_0^2\ll |\gamma_0|\ll \rho_0$ is met at  the ultraviolet scale, disorder can enhance the transition temperature $T_c$ compared to the clean case. The key point is that the first term in the RG equation for $\gamma$, namely the disorder-induced term $2\rho \gamma$, initially dominates and leads to an increase of $|\gamma|$ that is stronger compared to the clean case, until the resistance reaches $\rho^*\sim \rho_0^2/|\gamma_0|$. Afterwards, the second term, $2\gamma^2/3$, takes over and controls the final stage of the transition to the superconducting state. The resulting transition temperature can be estimated as $T_c\sim \tau^{-1}\exp(-2y^*)$, with $y^*\approx 1/\rho_0$,  which is larger than $T_{c0}\sim\tau^{-1}\exp(-1/|\gamma_0|)$ for the clean case. A more detailed analysis is provided in Ref.~\cite{Burmistrov15}.

\noindent
N.32\hspace{0.5cm}Still, many samples are needed in order to identify the universal properties of the MIT in $2d$. 

\noindent
N.33\hspace{0.5cm}$2d$ electron systems with nonzero spin or valley \emph{splitting} may include Si-MOSFET \cite{Simonian,Vitkalov,PudalovBauer} and $n$-AlAs-quantum wells \cite{Gunawan2006,Gunawan2007} as well as Al$x$Ga1$-x$As/GaAs/Al$x$Ga1$-x$As double quantum well heterostructures \cite{Minkov}. A discussion of such systems can be found in Ref.~\cite{Burmistrov2017}.

\noindent
N.34\hspace{0.5cm}This is, actually, a crucial point. The measurements in Ref.~\cite{Kuntsevich} confirmed the original idea \cite{AFPprl} that the difference between high- and low- mobility Si-MOSFET samples is in the strength of the inter-valley scattering rather than in $r_{s}$ itself, which even in the best samples is not too large. It was shown that in the Si-MOSFET devices the ratio $\tau_v/\tau_{el}$ monotonically increases when the electron density decreases; here $\tau_v$ is the time of the inter-valley scattering. On the other hand, electrons are always localized at low enough densities.  The peculiarity of the Si-MOSFETs fabricated from \emph{high-mobility} samples is that the MIT in the dilute electron liquid can be reached under conditions when the two valleys are distinct. For a review of low-mobility Si-MOSFETs, where the glass behavior was observed, see Ref.~\cite{Dragana2016}.

\noindent
N.35\hspace{0.5cm}The $g_{L}$-factor is about $1.5$ times larger than for free-electrons, i.e., $g_{L}/g_{L}^{0}=\frac{1}{1+F_{0}^{\sigma }}=1-\Gamma
_{\sigma }\approx 1.5$. The effective mass is about $3$ times larger than the band mass, $m^*/m_{b}\approx 3$. 

\noindent
N.36\hspace{0.5cm}The existence of the diffusive region is not obvious even in the neighborhood of the MIT. In \emph{high-mobility systems} like GaAs/AlGaAs or
n-SiGe heterostructures the single particle scattering rate typically differs by a factor of $10$ compared to the transport scattering rate. The smooth disorder drives the system directly from the ballistic to
the insulating phase. Therefore, it is difficult to access the diffusive regime.

\noindent
N.37\hspace{0.5cm}The problematic feature of the flow equations given by Eqs.~(\ref{onelooprho1}) and (\ref{oneloopgam1}) is that the amplitude $\gamma _{2}$
diverges at a finite temperature $T^{\ast }$ and thereafter the RG-calculation becomes uncontrolled~\cite{AF1984,CastellaniSpin}. 

(i) Fortunately, the scale $T^{\ast }$ decreases very rapidly with $n_{v}$ as $\ln \ln (1/\tau T^{\ast })\sim (2n_{v})^{2}$, making the problem of the divergence of $\gamma _{2}$ irrelevant for all practical purposes even for $n_{v}=2$ which corresponds to Si-MOSFETs \cite{AFPprl}. At $n_{v}\rightarrow \infty $, the theory becomes
internally consistent: $T^{\ast}\rightarrow 0$. 

(ii) For discussions of different scenarios related to the divergence of $\gamma _{2}$ at a finite $n_v$, see Refs. \cite{AF1984spin,AF1987ESR,BelitzK,CastDiCLeemr,ChM,NAleinerL}. 

\noindent
N.38\hspace{0.5cm} The validity of the WFL is closely connected with the quasiparticle description~\cite{Chester61,Langer62}. Experimental tests of the WFL through simultaneous measurements of electric and thermal transport are, therefore, an important tool for the detection of non-Fermi liquid behavior~\cite{Mahajan13}. A strong violation of the WFL was observed in graphene near the charge neutrality point where none of the assumptions underlying Fermi liquid theory are applicable \cite{Fong}.

\noindent
N.39\hspace{0.5cm} Of course, when only the linear response is studied, the temperature can \emph{eventually} be taken to be constant. However, this can be done only when the final expressions have already been obtained, rather than during the process of their derivation.

\noindent
N.40\hspace{0.5cm} The Cooperons and the $e$-$e$ interaction in the Cooper channel will be fully ignored. It is assumed that in the Cooper channel the $e$-$e$ interaction is repulsive, $\Gamma_c>0$. 

\noindent
N.41\hspace{0.5cm}  The corrections emerging from the sub-temperature energy range are generated by the same time-dependent part of the Coulomb interaction which is responsible for log-squared corrections to the TDOS. However, the argument that a purely time-dependent $e$-$e$ interaction can be eliminated by means
of a time-dependent gauge transformation which has been used for the electric conductivity is not applicable here, because this transformation gets entangled with the frequency-vertices of the correlation function $\chi_{kk}^{dyn}$.

\section*{Acknowledgements}
We are grateful to K.~Michaeli and A.~Punnoose for collaborations on topics discussed in this review. We have profited from discussions with many colleagues, but especially we wish to thank D.~E.~Khmelnitskii. This work was supported by the National Science Foundation under Grant No.~DMR-1742752 (G.S.).

\section*{Personal Note}
One of us (GS) was a PhD student of Konstantin Efetov’s. The other one (AF) was his long-time colleague at the Landau Institute for the Theoretical Physics. Konstantin’s pioneering work \cite{ELK} opened the perspective to generalize the NLSM by including $e$-$e$ interactions (Finkel'stein's model). Later, Konstantin introduced Georg to the physics of disordered electrons, and recommended him to continue his postdoctoral studies at the Weizmann Institute of Science, where the authors' collaboration began. This paper is our tribute to the memory of Konstantin Efetov, a teacher, a colleague and an outstanding scientist.

\end{document}